\documentclass[journal]{IEEEtran}
\usepackage{enumitem}
\usepackage{amssymb}
\usepackage{amsmath}
\usepackage{mathtools}
\usepackage{ellipsis}
\usepackage{textcomp}
\usepackage{cite}
\usepackage[pdftex]{graphicx}
\usepackage{epstopdf}
\usepackage{psfrag}
\usepackage{float}
\usepackage{grffile}
\usepackage{csquotes}
\usepackage{multicol}
\usepackage{subfigure}
\usepackage{epsfig}
\usepackage{multirow}
\usepackage{footnote}
\makesavenoteenv{tabular}
\makesavenoteenv{table}
\usepackage{booktabs, caption, makecell}

\usepackage{threeparttable}
\usepackage{tablefootnote}
\usepackage{pifont}
\usepackage{url}
\usepackage[normalem]{ulem}
\usepackage{color}
\usepackage{xcolor}

\usepackage{setspace}
\usepackage{amsthm}
\usepackage{array}
\usepackage{orcidlink}

\UseRawInputEncoding

\begin{document}
\title{UNet: A Generic and Reliable Multi-UAV Communication and Networking System Architecture for Heterogeneous Applications}

\author{Sanku~Kumar~Roy\orcidlink{0000-0003-0927-0130},~\IEEEmembership{Student~Member,~IEEE,}
Mohamed~Samshad\orcidlink{0009-0007-0298-4831},
and~Ketan~Rajawat\orcidlink{0000-0002-4508-0062},~\IEEEmembership{Member,~IEEE}
    \thanks{Manuscript received XYZ; revised XYZ.}
    \thanks{S.K. Roy was affiliated with the Department of Electrical Engineering at the Indian Institute of Technology, Kanpur, India, 208016, during this work. He is currently with the Department of Computing Science at the University of Alberta, Edmonton, AB, Canada, T6G 2E8 (e-mail: \texttt{sanku@ualberta.ca}).}
    \thanks{M. Samshad was affiliated with the Department of Electrical Engineering, Indian Institute of Technology, Kanpur, India, 208016, during this work. He is currently a Research Engineer with the National University of Singapore, Singapore (e-mail: \texttt{samshad@nus.edu.sg}).}
    \thanks{K. Rajawat is with the Department of Electrical Engineering, Indian Institute of Technology, Kanpur, India, 208016 (e-mail: \texttt{ketan@iitk.ac.in}).}
    \thanks{Digital Object Identifier XYZ}
}

\markboth{IEEE Transactions on Network and Service Management ,~Vol.~XYZ, No.~XYZ, XYZ~2026}%
{Shell \MakeLowercase{Roy \textit{et al.}}: UNet: A Generic and Reliable Multi-UAV Communication and Networking Architecture for Heterogeneous Applications}
\maketitle
\begin{abstract} 
The rapid growth of UAV applications necessitates a robust communication and networking system architecture capable of addressing the diverse requirements of various applications concurrently, rather than relying on application-specific solutions. This paper proposes a generic and reliable multi-UAV communication and networking system architecture designed to support the varying demands of heterogeneous applications, including short-range and long-range communication, star and mesh topologies, different data rates, and multiple wireless standards. Our architecture is designed for both ad hoc and infrastructure networks, ensuring seamless connectivity throughout the network. Additionally, we present the design of a multi-protocol UAV gateway that enables interoperability among various communication protocols to enhance connectivity. Furthermore, we introduce a data processing and service layer framework with a graphical user interface of a ground control station that facilitates remote control and monitoring from any location at any time. We practically implemented the proposed architecture and evaluated its performance using different metrics, demonstrating its effectiveness.
\end{abstract}
\section{Introduction}\label{prob2:sec:introduction}
\IEEEPARstart{U}{nmanned Aerial Vehicle Networks} (UAVNs) consist of unmanned aerial vehicles (UAVs) equipped with actuators, sensors, autopilot, and wireless communication, enabling them to share data with each other and a ground control station (GCS) \cite{bine2024}. UAVs can autonomously perform tasks without human intervention, offering flexibility, low operating costs, and easy deployment. These advantages make UAVs suitable for diverse applications in healthcare, agriculture, military, and more, including courier services, border security, disaster management, precision agriculture, and aerial photography \cite{kim2024cooperative, rahbari2024, Hentati2020}. However, reliable communication between UAVs and the GCS is essential for monitoring, controlling, and executing tasks successfully in real-time. The demands of these UAV applications vary in terms of communication distance, end-to-end delay, topology (e.g., point-to-point (P2P), point-to-multipoint (P2M), mesh), network size, and bandwidth. Some applications require a single UAV, while others may need multiple. Additionally, some cases need only remote monitoring with pre-instructed paths, while others demand real-time monitoring and control. These diverse requirements indicate the importance of designing a robust UAV communication and networking architecture for heterogeneous applications. This design raises key questions: \textit{how to support various networking demands, wireless protocols, real-time data service to users, and multiple applications concurrently.}

To address the problem mentioned above, existing works \cite{Venkatesh2017, zhang2019, Singh2018, Krichen2018, Jawhar2017, Hong2016} proposed UAV communication and networking architectures based on either single UAV or multi-UAV setups for specific applications. Venkatesh \textit{et al.} \cite{Venkatesh2017} proposed a fully autonomous UAV communication architecture using the TS832 serial wireless module, focusing on video transmission over P2P communication. However, this architecture is only suitable for single UAV applications. To address multi-UAV needs, Singh \textit{et al.} \cite{Singh2018} and Hong \textit{et al.} \cite{Hong2016} proposed multi-UAV communication architectures. Singh \textit{et al.} \cite{Singh2018} suggested a relay network of UAVs for video transmission to GCS using the Edimax WAP module, but it supports only short-range communication and lacks dynamic routing and mesh topology. On the other hand, Hong \textit{et al.} \cite{Hong2016} presented an LTE-based UAV communication architecture with P2P and P2M communication, but it does not support mesh topology and always depends on infrastructure networking. To overcome these issues, Krichen \textit{et al.} \cite{Krichen2018} and Jawhar \textit{et al.} \cite{Jawhar2017} introduced an abstract concept of multi-UAV communication architecture using different wireless standards, but without the in-depth analysis needed to address the challenges in designing a generic UAV communication and networking architecture. Unfortunately, they did not provide any solutions for overcoming these design challenges.

Existing works mostly focused on specific applications using a single wireless protocol and topology, limiting their applicability to homogeneous applications. These architectures are often incompatible with both infrastructure and infrastructure-less (ad hoc) networks. Cellular networks (e.g., 3G, 4G, 5G) offer reliable connectivity in urban areas but face coverage issues in rural, forest, and border regions and lack support for mesh topology, which is crucial for UAV swarms. In contrast, infrastructure-less networks support long-range, mesh, and star topologies with moderate data rates, allowing network expansion based on application needs. However, these networks are confined to specific regions where they are established. Therefore, there is a need for a generic UAV communication architecture that provides seamless connectivity between UAVs and GCS while concurrently supporting heterogeneous applications.

The trend of using IP-based wireless solutions allows UAVs to connect to the Internet for remote monitoring and control, making UAV data accessible from anywhere. However, existing architectures \cite{Li2013, Singh2018, Venkatesh2017, Hayat2015, Chriki2019} limit GCS accessibility to local areas, restricting global data service. Some of them \cite{Jawhar2017, Hong2016, Wang2018} have proposed multi-UAV communication architectures with global GCS accessibility, but they lack support for concurrent applications and heterogeneous wireless protocols. Thus, there is a need for a generic, reliable multi-UAV communication architecture that supports heterogeneous applications and wireless standards while ensuring global GCS accessibility. Note that, in this paper, the term ``generic" denotes that the proposed architecture is designed to support the diverse requirements of the heterogeneous applications discussed earlier.

In summary, our proposed architecture addresses the following key challenges in the paper:
\begin{itemize}
    \item Heterogeneous application requirements (distance, delay, topology, and bandwidth) and support for both real-time and non-real-time operations.
    \item Support for both single-UAV and multi-UAV systems, along with scalability in UAV networks and concurrent application support.
    \item Integration of multiple network topologies (P2P, P2M, mesh) and interoperability between infrastructure and ad hoc networks.
    \item Limitations of existing works, including dependence on single wireless protocols and lack of support for heterogeneous wireless standards.
    \item Coverage limitations of infrastructure networks and geographical confinement of ad hoc networks.
    \item Lack of global accessibility of GCS.
\end{itemize}

\noindent To address those key challenges, the main contributions of this paper are as follows:
\begin{itemize}[leftmargin=*, labelsep=0.5em]
\item We propose a generic UAV communication and networking system architecture, \textbf{UNet}, that supports heterogeneous applications concurrently. UNet is designed to meet the various application demands, including short-range and long-range communications, star and mesh topologies, various data rates, and heterogeneous wireless standards.
\item To ensure seamless connectivity between UAVs and GCS, we integrate support for both ad hoc and infrastructure networks and optimize key network parameters in the wireless modules.
\item Additionally, we propose the design of a multi-protocol UAV gateway to enable interoperability among UAVs utilizing different communication protocols.
\item We introduce a data processing and service framework integrated with the GCS to enable remote control and monitoring from anywhere, at any time.
\item We validate the proposed architecture through practical implementation and comprehensive evaluation in outdoor environments using performance metrics selected to capture the requirements of heterogeneous UAV applications, rather than focusing on individual application scenarios.
\end{itemize}

The remainder of this paper is organized as follows: Section~\ref{sec_related_works} reviews the state-of-the-art UAV communication and networking architectures, highlighting their limitations. Section~\ref{sec_system_model} describes the problem scenario and introduces the novel UNet architecture. Section~\ref{sec_design_UNet} presents the design and operational flow of UNet. Section~\ref{sec_implementation_UNet} discusses the implementation of UNet. Section~\ref{sec_performance_evalution} presents the experimental setup and evaluates the performance of UNet. Finally, we conclude our work by discussing possible future research directions in Section~\ref{sec_conclusion_future_work}.
\begin{table*}[!ht]
        \centering
	\caption{Comparison of UNet with State-of-art existing works}
	\label{table_comparison}
	\resizebox{\textwidth}{!}{%
	\begin{tabular}{|l|l|l|l|l|l|l|l|l|l|}
		\hline
		\textbf{Author}          & \textbf{Muti-UAVs} & \textbf{Ad hoc Mesh} & \textbf{U2I} & \textbf{Range} & \textbf{Scalability} & \textbf{Generic Arch} & \textbf{Imple} & \textbf{GCS} & \textbf{Remarks} \\ \hline
		UNet            &  Yes                  & Yes            &     Yes         &          both short and long      &     Yes              &             Yes          &       Yes                  &       Globally           &	Prototyping\\ \hline
		Li \textit{et al.}\cite{Li2013}            &  Yes                  & Yes            &     No         &          --      &     Yes              &             No          &       No                  &       Locally           &	Abstract idea\\ \hline
		Singh \textit{et al.}\cite{Singh2018}            &  Yes                  & No full mesh            &     No         &          Short      &     No              &             No          &       Lab scale                  &       Locally           &	\\ \hline
		Venkateshet \textit{et al.}\cite{Venkatesh2017}            &     No               & No             &   No           &      Short          &     No              &          No             &            Yes             &   Locally               & 		\\ \hline
		Hayat \textit{et al.}\cite{Hayat2015}    &    Yes                & Yes                &       No       &        Short        &         No          &    No                   &        Yes                 &    Locally              & 		\\ \hline
		Krichen \textit{et al.}\cite{Krichen2018}            &       Yes             & Yes                   &        Yes      &       --         &    --               &           --            &               No          &       --           & Abstract idea\\ \hline
		Jawhar \textit{et al.}\cite{Jawhar2017}     &  Yes                  &      Yes            &       Yes       &    Long            &       Yes            &            No           &        Yes                 &         Globally         & Just idea\\ \hline
		Chriki \textit{et al.}\cite{Chriki2019}
		& Yes                   &    No            &     No         &    --            &        No           &         No              &   No                      &      Locally            & \\ \hline
        Silva \textit{et al.}\cite{silva2017}
		& Yes                   &    Yes            &     No         &    Short            &        No           &         No              &   Yes                      &      Locally            & \\ \hline
		Hong \textit{et al.}\cite{Hong2016}
		& Yes                   &    No            &     Yes         &    Long           &        Yes           &         No              &   No                      &      Globally            & \begin{tabular}[c]{@{}l@{}}Always Infrastructure\\-dependent\end{tabular}\\ \hline
		Wang \textit{et al.}\cite{Wang2018}
		& Yes                   &    --            &     Yes         &    Long           &        Yes           &         No              &   No                      &      Globally            & \\ \hline
	\end{tabular}
	}
\end{table*}
\section{Related Works} \label{sec_related_works}
Table \ref{table_comparison} summarizes related work and compares the UNet with existing UAV communication architectures, highlighting the key features needed to design a versatile system for heterogeneous applications. Below, we discuss each work and its limitations.

For single UAVs, researchers have proposed P2P architectures, while P2M and mesh architectures are discussed for multiple UAVs. Venkatesh \textit{et al.} \cite{Venkatesh2017} introduced a fully autonomous P2P communication architecture for video transfer to GCS using the TS832 wireless module. However, this is unsuitable for multiple UAVs, long-range, or IP networks. Multi-UAV communication architectures have been explored by others \cite{Chriki2019, Hong2016, Li2013, Singh2018, Hayat2015, Wang2018, silva2017}. For example, Singh \textit{et al.} \cite{Singh2018} proposed a relay-based architecture using the EDIMAX WAP module, where each module manually selects its next-hop routing. This is unreliable due to node dependency and a lack of support for mesh topology and dynamic routing. Chriki \textit{et al.} \cite{Chriki2019} presented a star topology for bandwidth sharing, with GCS as the central node, but it does not support swarm networking or specify the wireless protocol. Similarly, Hong \textit{et al.} \cite{Hong2016} proposed an LTE-based architecture for control and non-payload communication, but it lacks mesh networking and relies on LTE infrastructure. Silva \textit{et al.} \cite{silva2017}, and Indu \textit{et al.}\cite{indu2024} proposed a multi-UAV architecture for scanning rocket impact areas and disaster response missions, respectively. In \cite{silva2017}, the architecture was designed based on XBee Pro 900HP S3B wireless modules, and on the other hand, \cite{indu2024} used the WiFi standard in their centralized and decentralized architectures. Their architecture application is limited to a specific application due to the usage of a single protocol and is not integrated with the infrastructure network.

Khan \textit{et al.} \cite{Khan2017} explored various UAV ad hoc networking architectures and suitable routing protocols for flying ad hoc networks (FANETs), like destination-sequenced distance vector, optimized link state routing, dynamic source routing, and ad hoc on-demand distance vector, but did not offer a comprehensive communication solution. Li \textit{et al.} \cite{Li2013} compared four architectures and introduced multi-layer UAV ad hoc networking, discussing data link layer protocols for military applications. Krichen \textit{et al.} \cite{Krichen2018} reviewed communication systems between UAVs and GCS, presenting a holistic UAV architecture that includes multi-UAVs, WSNs, cellular networks, and satellites. Jawhar \textit{et al.} \cite{Jawhar2017} outlined UAV system requirements, functions, and services, discussing U2U and U2I communication architectures and possible wireless standards. Both Krichen \textit{et al.} and Jawhar \textit{et al.} emphasized the need for multi-UAV architectures supporting heterogeneous applications without detailed solutions.

Zhang \textit{et al.} \cite{zhang2019} proposed a swarm network architecture with a low-latency routing algorithm, but it is application-specific, supports only ad hoc networking, and does not address remote GCS data transmission or infrastructure integration. An adaptive architecture proposed in \cite{huang2018} selects between mobile ad hoc network (MANET) and disruption tolerant networking based on data characteristics, noting that MANET bandwidth decreases with more hops, but it fails to integrate infrastructure and ad hoc networks or address remote GCS.

\noindent\textit{\underline{Synthesis}}: Table \ref{table_comparison} summarizes the features of existing works and the missing features in their designed architectures to address the key challenges. Later, a critical review exactly pinpointed their limitations in detail, revealing a gap in designing a generic UAV communication architecture for concurrent heterogeneous UAV applications. Most studies focus on application-specific architectures \cite{Venkatesh2017, Chriki2019, Hong2016, Li2013, Singh2018, Hayat2015, Wang2018, silva2017, indu2024}, while a few discuss more generic architectures \cite{Krichen2018, Jawhar2017, zhang2019, huang2018} but lack detailed and feasible solutions from a practical deployment perspective. Furthermore, existing architectures often fail to address remote monitoring and control. This paper introduces a novel system architecture that supports multiple UAV applications, integrates ad hoc and infrastructure networks, and introduces middleware to handle multiple UAV networks concurrently, enabling remote monitoring and control through both hardware and software components.
\section{System Model} \label{sec_system_model}
In this section, we outline the problem scenario and introduce the proposed UNet architecture, detailing its layers and components for enabling efficient UAV communication and networking. 
\subsubsection{Problem Scenario}\label{prob2:ssec:problem-scenario}
\begin{figure}[t]
	\centering
	\includegraphics[width=9cm]{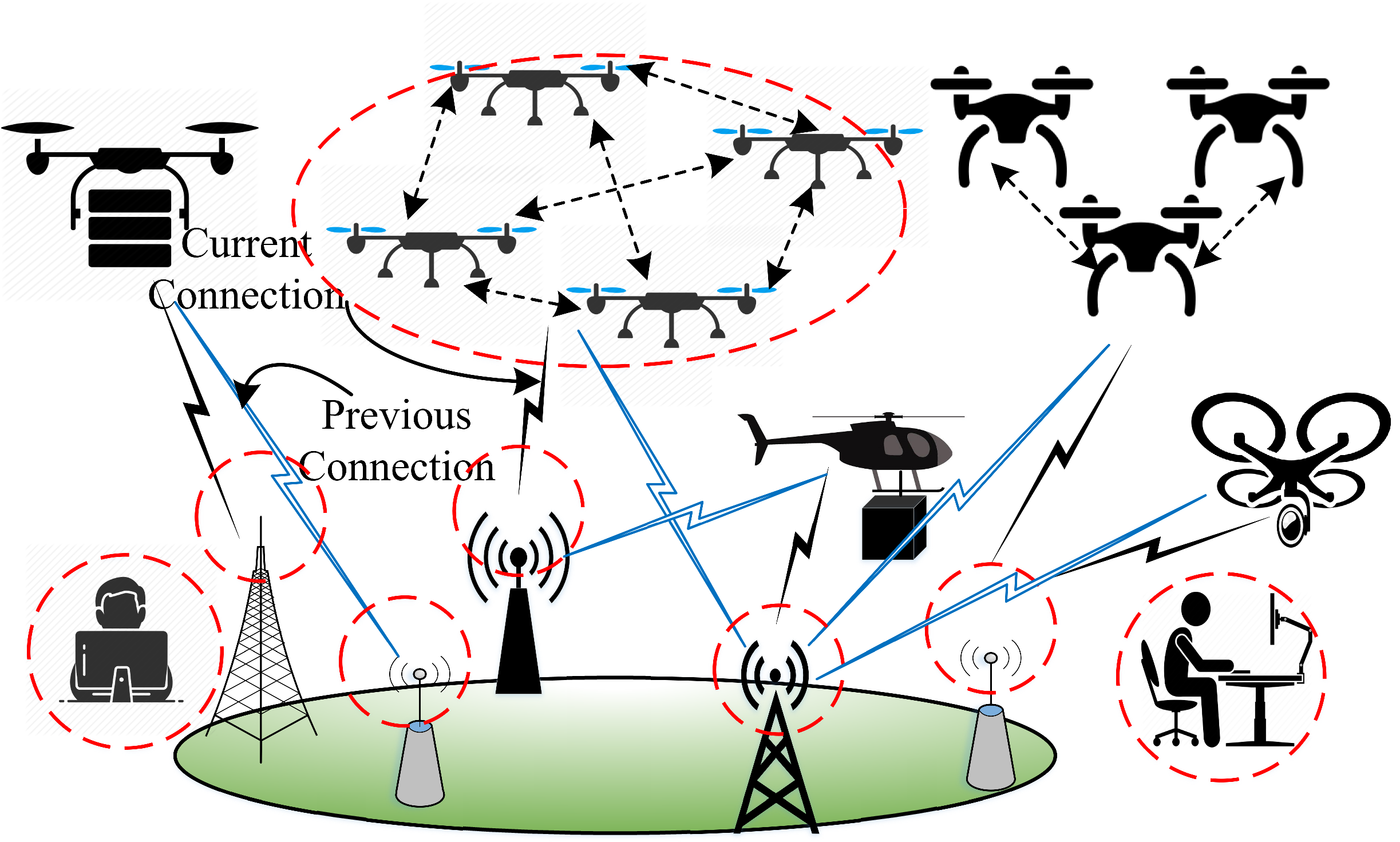}
	\caption{Problem Scenario}
	\label{fig_problem_scenario}
\end{figure}

Designing a generic communication and networking architecture for UAVs presents significant challenges, especially when supporting a wide range of applications, as shown in Fig. \ref{fig_problem_scenario}. To 
meet the objectives, the following key issues must be addressed:
\begin{itemize}[leftmargin=*, labelsep=0.5em]
\item Integrating heterogeneous wireless protocols into a unified architecture while accommodating various UAV applications,
\item Ensuring seamless connectivity between UAVs and GCS, even when UAVs operate at high speeds and over long distances,
\item Implementing robust mesh networks for UAV swarms,
\item Enabling globalized GCS for remote monitoring and control.
\end{itemize}

In Fig. \ref{fig_problem_scenario}, every UAV application uses a specific wireless protocol for communication between UAVs and GCS. However, when integrating multiple networks with heterogeneous applications into the same architecture, it becomes problematic due to each network's unique wireless protocol. The distinct data link and physical layers of these protocols make integrating heterogeneous networks with a UAV communication architecture more complex. Additionally, UAVs are highly mobile, covering long distances at certain altitudes to execute tasks. Maintaining wireless connectivity between UAVs and GCS during such missions is extremely challenging, especially during travel, leading to intermittent connectivity.

In some cases, UAV applications require a swarm of UAVs to perform specific tasks. Implementing mesh topology in such FANETs is challenging due to UAVs' high mobility and dynamic formations. In particular, selecting a routing protocol that accommodates frequent link quality changes, high mobility, and energy constraints is crucial. Furthermore, traditional UAV communication architectures are often proprietary and vendor-specific, making them inflexible for managing multiple applications remotely and delivering data to users. Thus, designing a remote GCS capable of supporting multiple applications concurrently presents a significant challenge.

\subsubsection{Architecture of UNet}
\begin{figure}[t]
	\centering
	\includegraphics[width=8.5cm]{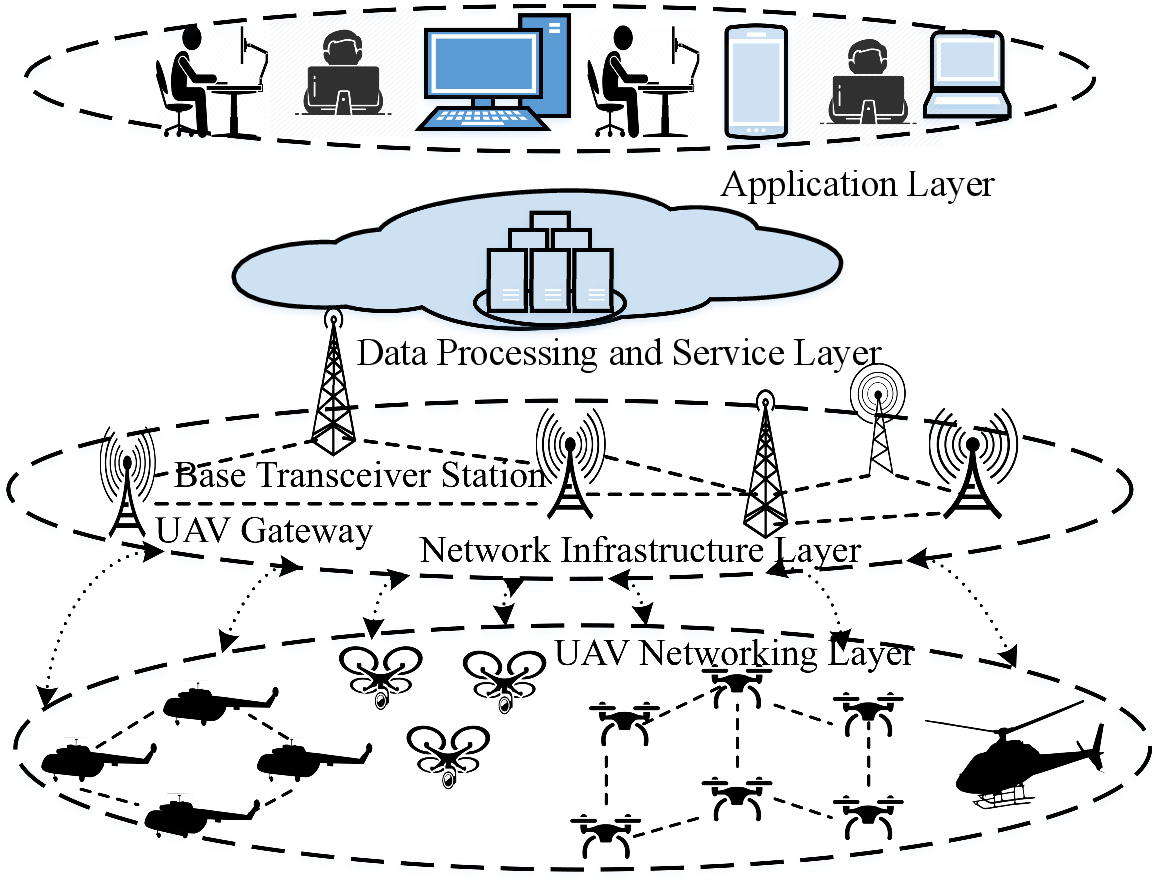}
	\caption{The proposed architecture}
	\label{fig_proposed_architecture}
\end{figure}

Fig. \ref{fig_proposed_architecture} presents the architecture of UNet, outlining its layers and components. UNet ensures end-to-end communication between UAVs and GCS by incorporating UAV-to-UAV (U2U), UAV-to-Infrastructure (U2I), and Infrastructure-to-Infrastructure (I2I) communications. Notably, the architecture does not rely on homogeneous or heterogeneous wireless protocols used by application networks. UNet is divided into four layers: a) UAV networking layer (UNL), b) UAV networking infrastructure layer (UNIL), c) data processing and service layer (DPSL), and d) application layer (AL).

\noindent\textit{i) UAV Networking Layer:}
UNL, the lowest layer, consists of UAVs communicating using P2P and multi-hop routing protocols. UAVs connect in ad hoc, infrastructure, or hybrid networks, depending on application needs. UNL supports various wireless protocols and topologies, including P2P, P2M, and mesh. In P2P networks, UAVs transmit data directly to UNIL via U2I communication. In mesh topologies, UAVs share data with each other using U2U links and then forward it to UNIL via U2I. In P2M networks, UAVs communicate with a central node using U2U, which relays the data to UNIL.

\noindent\textit{ii) UAV Networking Infrastructure Layer:}
UNIL bridges UNL and DPSL, forwarding UAV data to and from GCS. It ensures seamless wireless connectivity for various UAV networks, whether ad hoc or infrastructure-based, depending on the application's environment. UNIL includes UAV gateways for ad hoc networks and base transceiver stations (BTSs), which provide 2G to 5G connectivity via telecommunications infrastructure. All gateways and BTSs are connected to a core network, which forwards UAV data to the remote GCS in DPSL.

\noindent\textit{iii) Data Processing and Service Layer:}
DPSL processes incoming UAV data and provides services to remote users. It includes two units: 1) the data processing unit, which manages UAV communication, and 2) the GCS unit, which handles data services for remote users. The GCS enables web-based monitoring and control of UAVs and applications.

\noindent\textit{iv) Application Layer:}
This layer involves remote users who generate application-specific requests. These requests are processed by the GCS at DPSL, with acknowledgments returned to users. Through the GCS web service, users can monitor network activities and control their UAVs remotely.

\section{Design of UNet}\label{sec_design_UNet}
In this section, we describe the design of the UAV architecture, gateway, and data processing unit, focusing on their components, communication protocols, and the framework used to manage UAV data and services.
\begin{figure}[t]
	\centering
	\includegraphics[width=8cm]{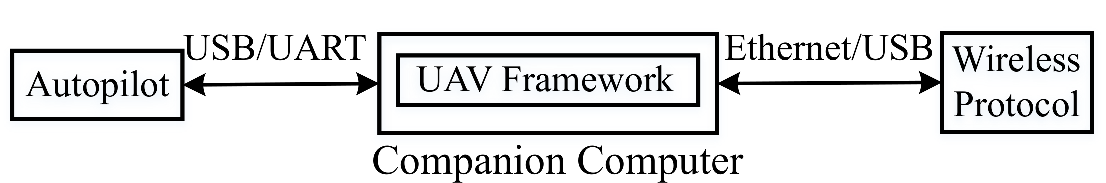}
	\caption{The block diagram of UAV architecture}
	\label{fig_UAV_architecture}
\end{figure}

\noindent\textit{1) Design of UAV Architecture:}
The design of a UAV architecture plays a crucial role in developing a UAV communication system. Fig. \ref{fig_UAV_architecture} shows the different components of the UAV, which include: 1) autopilot, 2) companion computer, and 3) wireless protocol. The autopilot, the core of the UAV, controls its trajectory without constant manual input. It combines flight control hardware with algorithms to stabilize UAV flight. Sensors like gyroscopes, magnetometers, barometers, GPS, and actuators such as motors and auto-landing gear are integrated into the autopilot. The autopilot connects to a companion computer to transfer telemetry data via interfaces like USB or UART.
\begin{figure}[h]
	\centering
	\includegraphics[width=8cm]{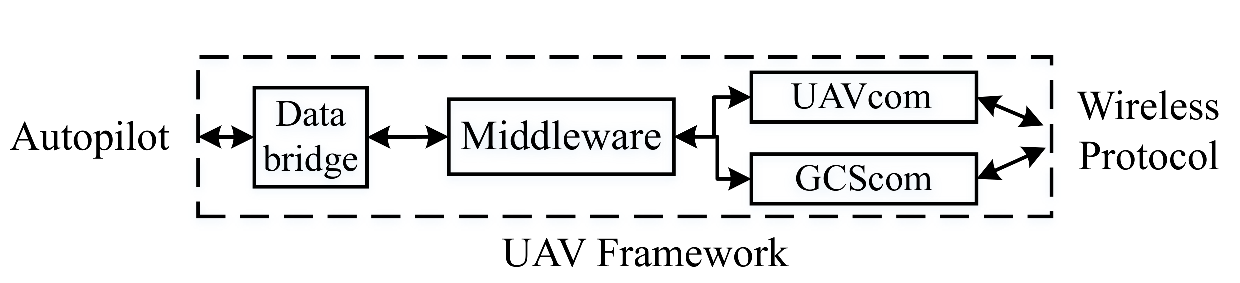}
	\caption{The block diagram of the UAV framework}
	\label{fig_UAV_node_framework}
\end{figure}

The companion computer reduces the data generated by the autopilot and shares it with neighboring UAVs and the DPSL. It also supports high-speed interfacing with wireless modules. The architecture follows a publisher-subscriber framework (PSF). Fig. \ref{fig_UAV_node_framework} outlines the UAV framework, which includes: 1) data bridge, 2) middleware, 3) UAVcom, and 4) GCScom. The data bridge converts communication between the autopilot and middleware. Middleware handles three types of messages: topic (sensor data, battery, diagnostics), service (commands like landing or takeoff), and acknowledgment messages. The framework collects sensor data from the autopilot through the data bridge, processes it, and shares it with UAVcom and GCScom, minimizing unnecessary data to reduce network traffic and latency. GCScom manages communication with the DPSL, while UAVcom handles UAV-to-UAV communication. The companion computer also supports swarm algorithms and various routing protocols for UAV networks.
\begin{figure}[!ht]
	\centering
	\includegraphics[width=5.5cm]{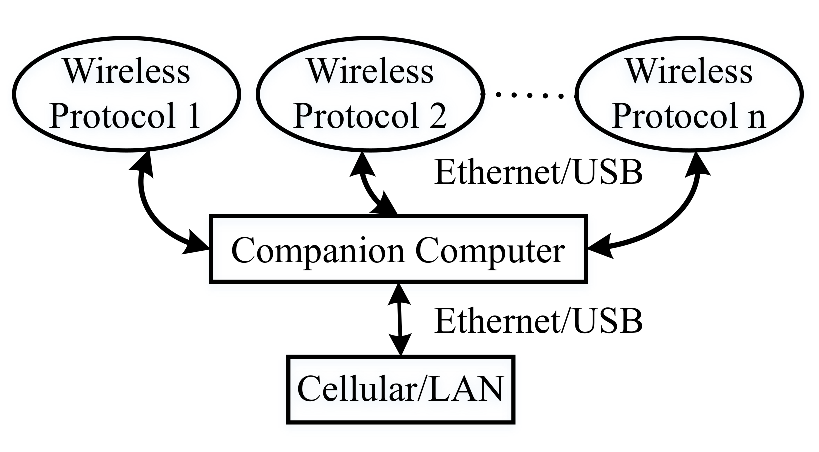}
	\caption{The block diagram of UAV gateway}
	\label{fig_gateway_architecture}
\end{figure}

The wireless protocol is critical for UAV communication. The wireless unit connects to the companion computer via Ethernet (IEEE 802.3) or USB. Network parameters like topology, data rate, delay, and range depend on the chosen wireless protocol, which influences network performance. UAV architecture can accommodate either ad hoc or infrastructure wireless protocols to generalize the communication architecture for heterogeneous protocols and support both long- and short-range communication.

\noindent\textit{2) Design of UAV Gateway:}
The UAV gateway supports multiple wireless networks using different protocols to serve heterogeneous UAV applications, as shown in Fig. \ref{fig_gateway_architecture}. It enables seamless communication between different wireless protocols. One side supports ad hoc protocols, while the other uses infrastructure protocols. The gateway facilitates a data flow between networks with heterogeneous wireless protocols and includes ad hoc wireless protocols, a companion computer, and infrastructure protocols. It collects data from UAVs via ad hoc protocols and transfers it to the DPSL via cellular or LAN. Cellular and LAN networks are connected to the Internet, extending the communication range between GCS and UAVs. Infrastructure protocol-based UAVs are directly connected to BTSs.
\begin{figure}[h]
	\centering
	\includegraphics[width=7.5cm]{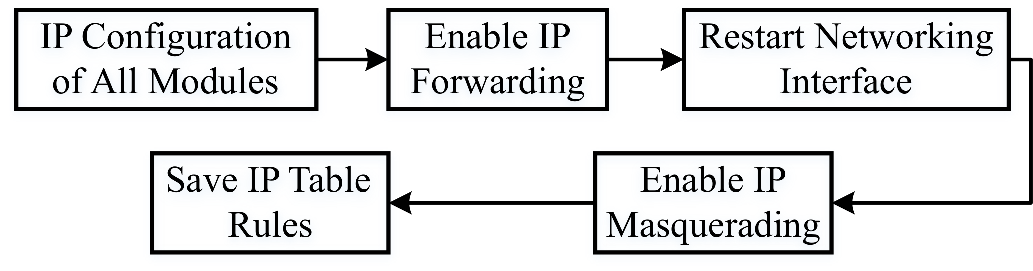}
	\caption{The procedure of UAV gateway configuration}
	\label{fig_gateway_configuration}
\end{figure}

Fig. \ref{fig_gateway_configuration} illustrates the gateway configuration process. First, configure the wireless modules connected to the companion computer by assigning a unique IP, gateway, subnet mask, and DNS. Next, IP forwarding is enabled in the companion computer's OS. Restart the networking interface and enable IP masquerading to set up the IP gateway. Finally, save all rules in the IP table.
\begin{figure}[h]
	\centering
	\includegraphics[width=9.5cm]{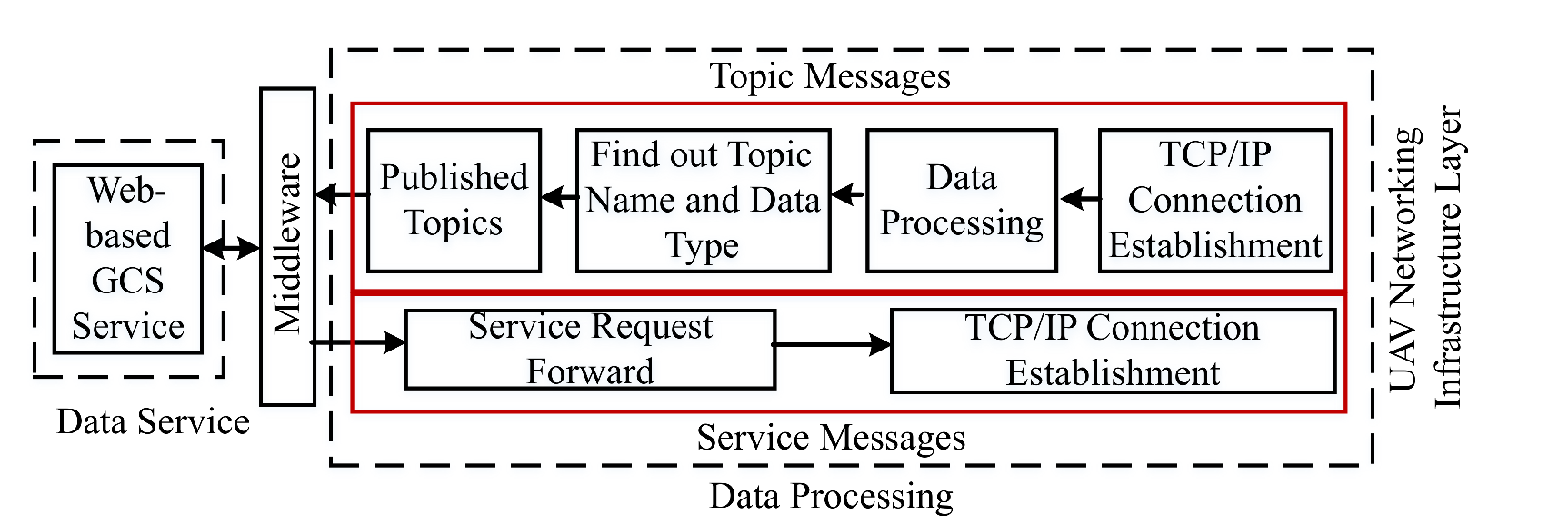}
	\caption{The block diagram of data processing and service unit}
	\label{fig_data_processing_service_layer}
\end{figure}
\begin{table*}[t]
\centering
	\begin{threeparttable}
		\scriptsize
		\caption{Specifications of the different wireless modules used}
		\label{table_wireless_protocol_specifications}
		\begin{tabular}{|c|c|c|c|c|c|}
			\hline
			\textbf{Wireless Modem} & \textbf{Standard} & \textbf{Distance (km)} & \textbf{Topology}   & \textbf{Data rate (Mbps)} & \textbf{Interface Protocol} \\ \hline
			AlfaTube 2H\cite{alfanetwork2019}            & IEEE802.11b/g/n   & Long Range                       & Mesh\tnote{*}, P2M, and P2P &       150                    & Ethernet                    \\ \hline
			Ubiquiti Bullet M2\cite{bullet2019}      & IEEE802.11b/g/n   & 50+                    & Mesh\tnote{*}, P2M, and P2P & 100                       & Ethernet                    \\ \hline
			TP-Link WR902AC\cite{tplink2019}          & IEEE802.11n/ac    & 0.433                    & Mesh\tnote{*}, P2M, and P2P &       300                    & Ethernet                    \\ \hline
			Microhard pMDDL2450 \cite{microhard2019}     & Private           & 20                     & Mesh, P2M, and P2P  & 25                        & Ethernet                    \\ \hline
			JioFi JMR540\cite{jio2019}              & 4G                & Depends on GSM         & P2P                 & 150                       & USB                         \\ \hline
		\end{tabular}
		\begin{tablenotes}\footnotesize
			\item[*] After configuring mesh with the help of OpenWrt\cite{openwrt2019}.
		\end{tablenotes}
	\end{threeparttable}
\end{table*}  

\noindent\textit{3) Design of Data Processing and Service Unit:}
The data processing and service unit has three main components: data processing, middleware framework, and data service. Fig. \ref{fig_data_processing_service_layer} shows the unit's functions. The middleware framework, a PSF, serves as an information bridge between data processing and services. The data processing unit handles incoming data (topic messages) from UNIL and outgoing services (service messages) requested by users from the application layer. It establishes a secure TCP/IP connection with UAVs to publish topic messages and extract meaningful information, which is then shared through the framework. A second secure TCP/IP connection is used to send service commands to UAVs.

Data service is a web-based GCS for remotely monitoring and controlling UAV activities. The GCS subscribes to the necessary topics from the PSF to display sensor information. When users send control instructions, the GCS requests a service command from the PSF, which forwards it to the data processing unit to transmit to UAVs.

In summary, the proposed architecture and its design discussion provide evidence of how each component contributes to achieving the main objective of this work. The design of the UAV ensures communication among all onboard components, and users can choose any IP-based wireless communication protocol based on their application requirements. The UAV gateways enable seamless communication with UAVs that use infrastructure-less networks, even when they use heterogeneous wireless protocols and move/fly between gateways. The corresponding handover delay is discussed in the results section. If any UAVs use infrastructure networks, they can directly communicate with the data processing and service layer via base transceiver stations. The data processing and service layer handles multiple applications simultaneously, irrespective of the nature of the applications, whether they are homogeneous or heterogeneous. The layer processes incoming and outgoing UAV data in the form of ROS topics and services, and provides services to the remote users for remote controlling and monitoring their UAV activities from the application layer. At the application layer, the GCS Web interface displays ongoing application information, including sensor readings, and provides a convenient interface for sending service instructions to control UAVs.
\section{Implementation of {UNet}} \label{sec_implementation_UNet}
In this section, we detail the implementation of the UAV architecture, focusing on the integration of autopilot, UAV framework, wireless protocols, and the deployment of the UAV gateway and data processing 
layer.
\subsection{Implementation of UAV Architecture}
\noindent\textit{1) Implementation of Autopilot:}
In the previous section, we discussed the three main components of our designed UAV architecture: autopilot, UAV framework, and wireless protocol (Fig. \ref{fig_UAV_architecture}). We used the Pixhawk2 \cite{pixhawk2018} as the autopilot hardware along with the ArduPilot firmware stack, which generates sensor data, actuator data, and control messages in the MAVLink format and sends them to the companion computer via USB. We use either the UP squared ($UP^2$) developer board \cite{upboard2019} or Raspberry Pi 4 Model B \cite{raspberrypi2019} as the companion computer, with Ubuntu 20.04 LTS as the operating system.

\begin{figure}[h]
	\centering
	\includegraphics[width=6.5cm]{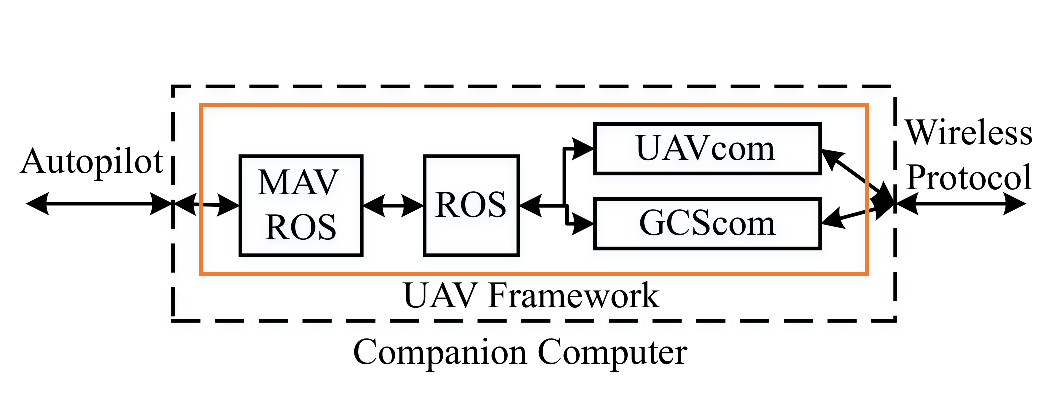}
	\caption{The block diagram of UAV framework implementation}
	\label{fig_UAV_framework_process}
\end{figure}

\noindent\textit{2) Implementation of UAV Framework:}
The step-by-step process of the UAV framework is shown in Fig. \ref{fig_UAV_node_framework}, with tools/software detailed in Fig. \ref{fig_UAV_framework_process}. The framework includes four main units: robot operating system (ROS) \cite{ros2019}, MAVROS, UAVcom, and GCScom, all of which are software layers. ROS \cite{ros2019} serves as a middleware designed to support heterogeneous computer clusters, using ROS topics and ROS services for data. MAVROS acts as a bridge between Pixhawk2 and ROS, converting MAVLink-based data from the Pixhawk2 into ROS topics and services, and vice versa. The ROS block publishes these topics and executes the services.

\begin{figure}[!ht]
	\centering
	\hspace{-3mm}
	\subfigure[DPSL side]
	{
		\includegraphics[width=3.9cm]{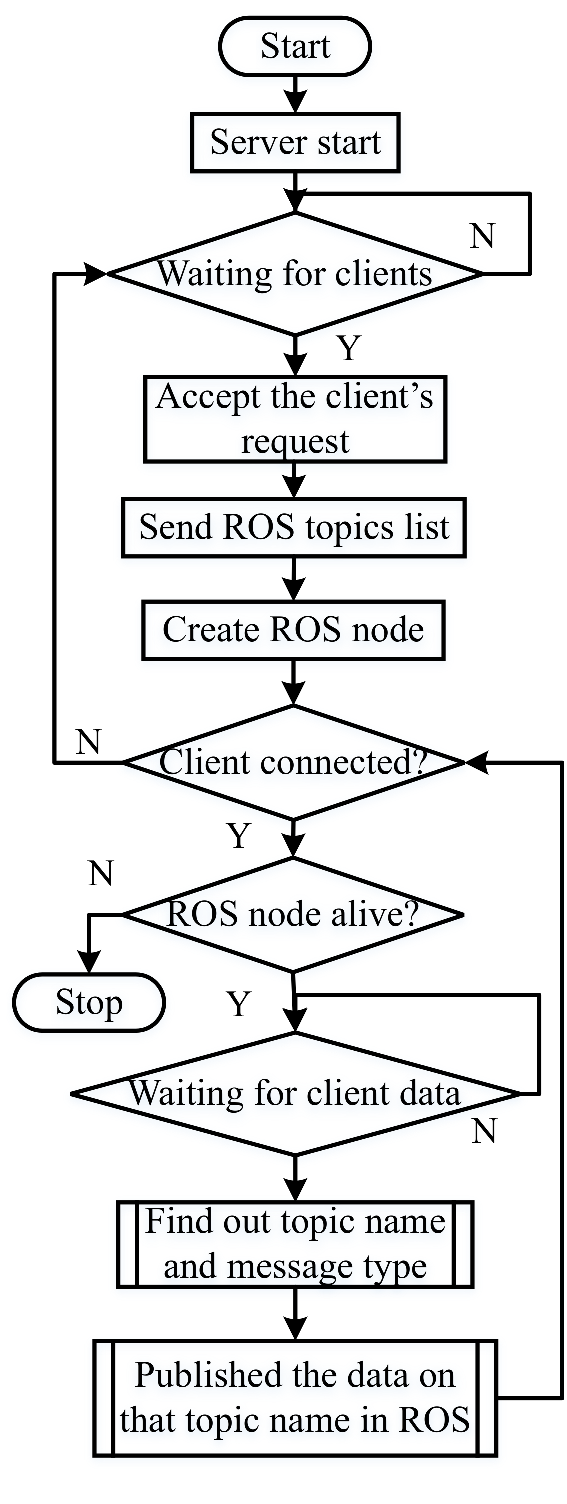}
		\label{fig_DPSL_server_rostopics}	
	}
	\subfigure[UAV side]
	{
		\includegraphics[width=3.9cm]{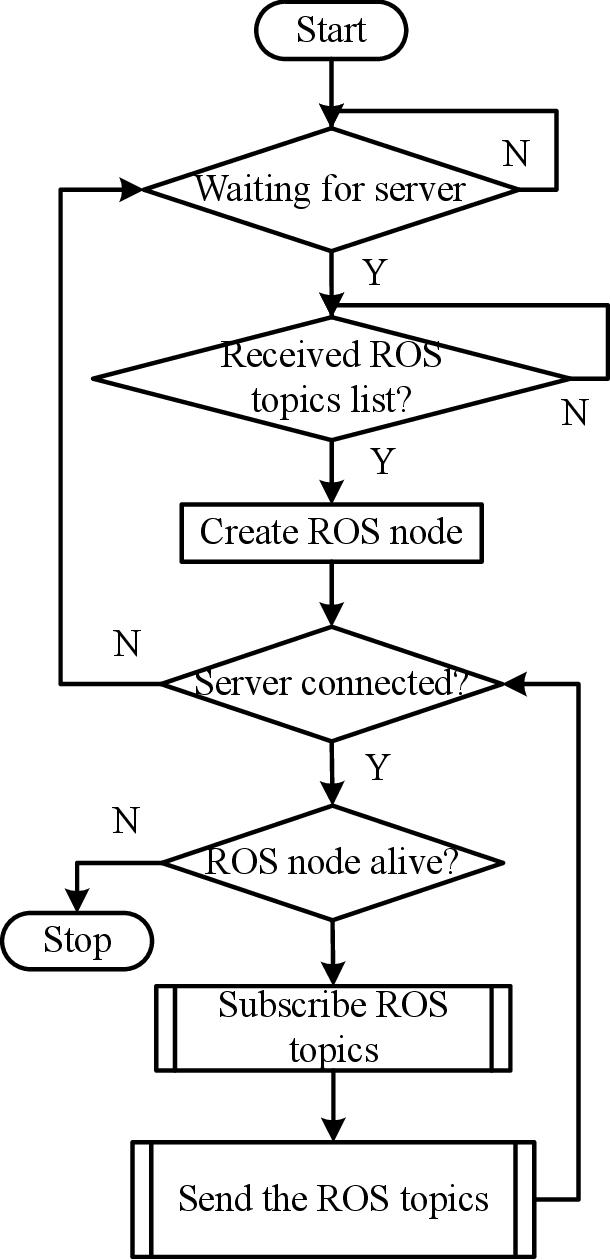}
		\label{fig_DPSL_client_rostopics}	
	}
	\caption{The flowchart of ROS topics handling between UAV and DPSL (a) DPSL side (b) UAV side.}
	\label{fig_DPSL_rostopics}
\end{figure}

\begin{figure}[!ht]
	\centering
	\hspace{-3mm}
	\subfigure[DPSL side]
	{
		\includegraphics[width=3.9cm]{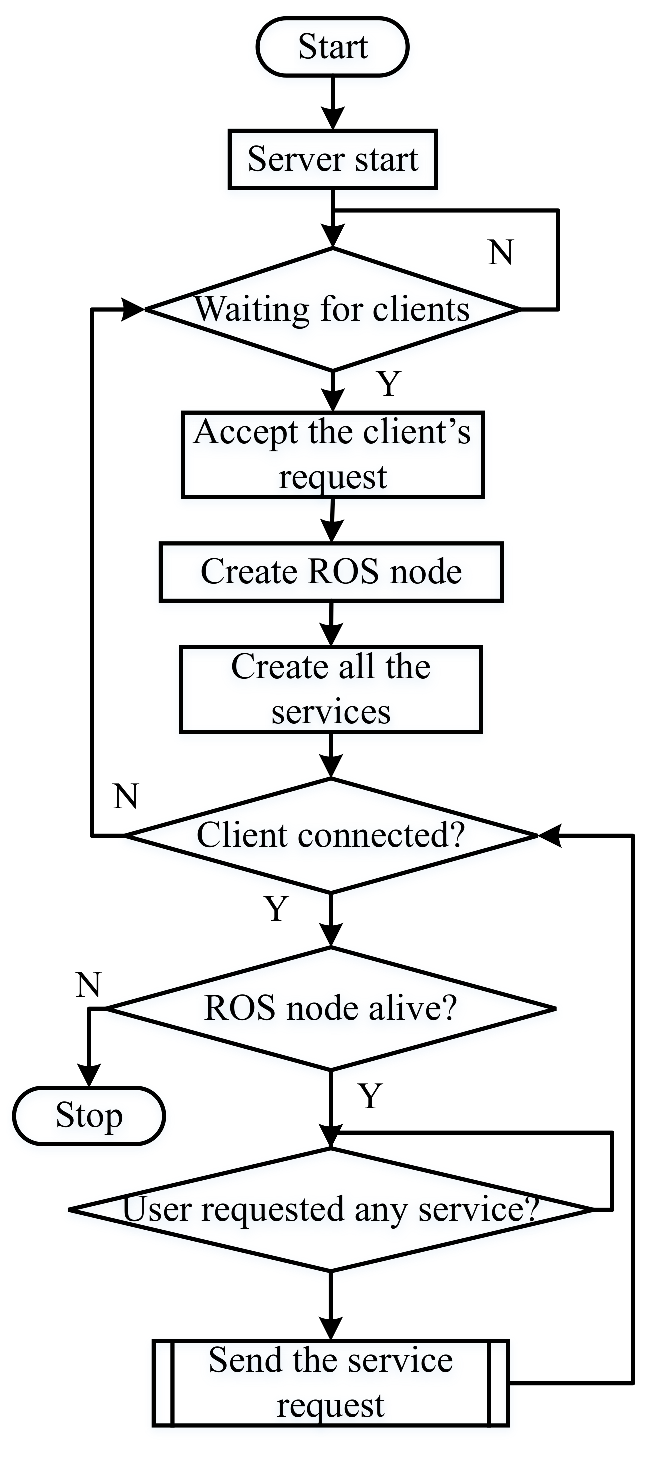}
		\label{fig_DPSL_server_service}	
	}
	\subfigure[UAV side]
	{
		\includegraphics[width=3.9cm]{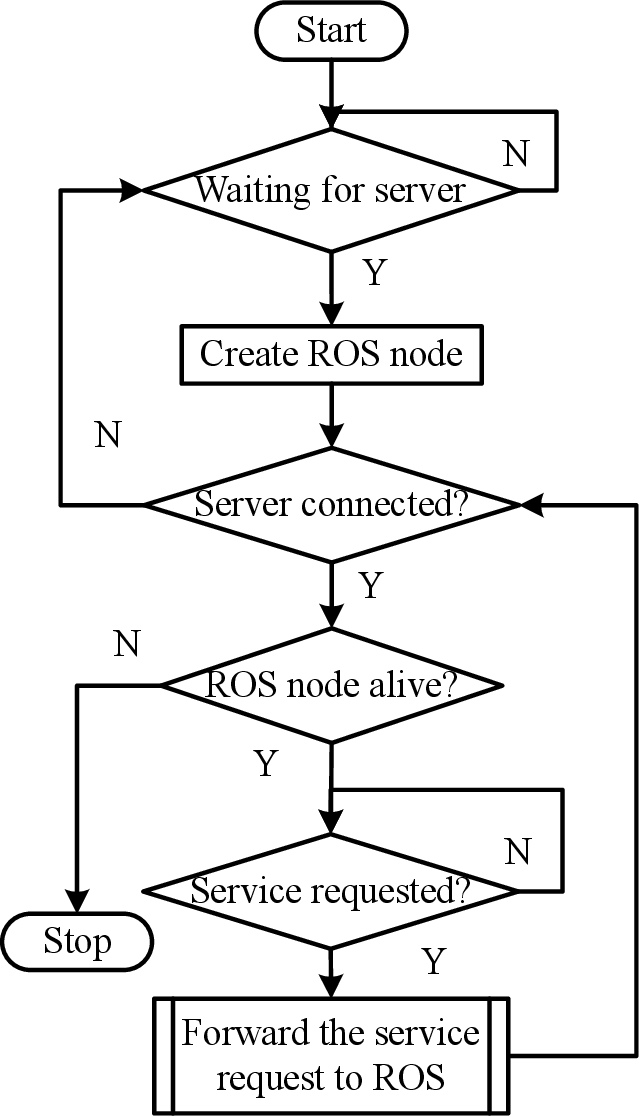}
		\label{fig_DPSL_client_service}	
	}
	\caption{The flowchart of ROS services handling between UAV and DPSL. (a) DPSL side. (b) UAV side}
	\label{fig_DPSL_rosservice}
\end{figure}

UAVcom and GCScom handle data communication between UAVs and GCS from the application and transport layers perspectives. UAVcom manages the sharing of ROS topics and services in P2P, P2M, or mesh topologies using the wireless protocol. In P2M, a single master and multiple slaves are configured, while in swarm formation, all UAVs function as masters, making the mesh more reliable. The `multimaster\_fkie' package is used to set up multi-UAV data communication on top of the different topologies. UAVcom also forwards topics and services from neighboring UAVs within the same network. GCScom maintains a virtual connection for data sharing between DPSL and the UAV’s ROS, using secure TCP/IP connections to share topics and services, as shown in Figs. \ref{fig_DPSL_client_rostopics} and \ref{fig_DPSL_client_service}.

\noindent\textit{3) Implementation of Wireless Protocol Unit:}
In UNet, we used various wireless modules, detailed in Table \ref{table_wireless_protocol_specifications}. Modules like AlfaTube 2H, Ubiquiti Bullet M2, and TP-Link WR902AC support P2M and P2P topologies but not mesh. However, we implemented mesh topology using OpenWrt \cite{openwrt2019} to enable swarm formation. These wireless modules meet the needs of diverse applications, including short-range with high data rates, long-range with moderate data rates, and long-range with high data rates, as well as different topology formations like P2P, P2M, and mesh.

\begin{figure}[!ht]
	\centering
	\hspace{-3mm}
	\subfigure[Quadcopter]
	{
		\includegraphics[width=4.2cm]{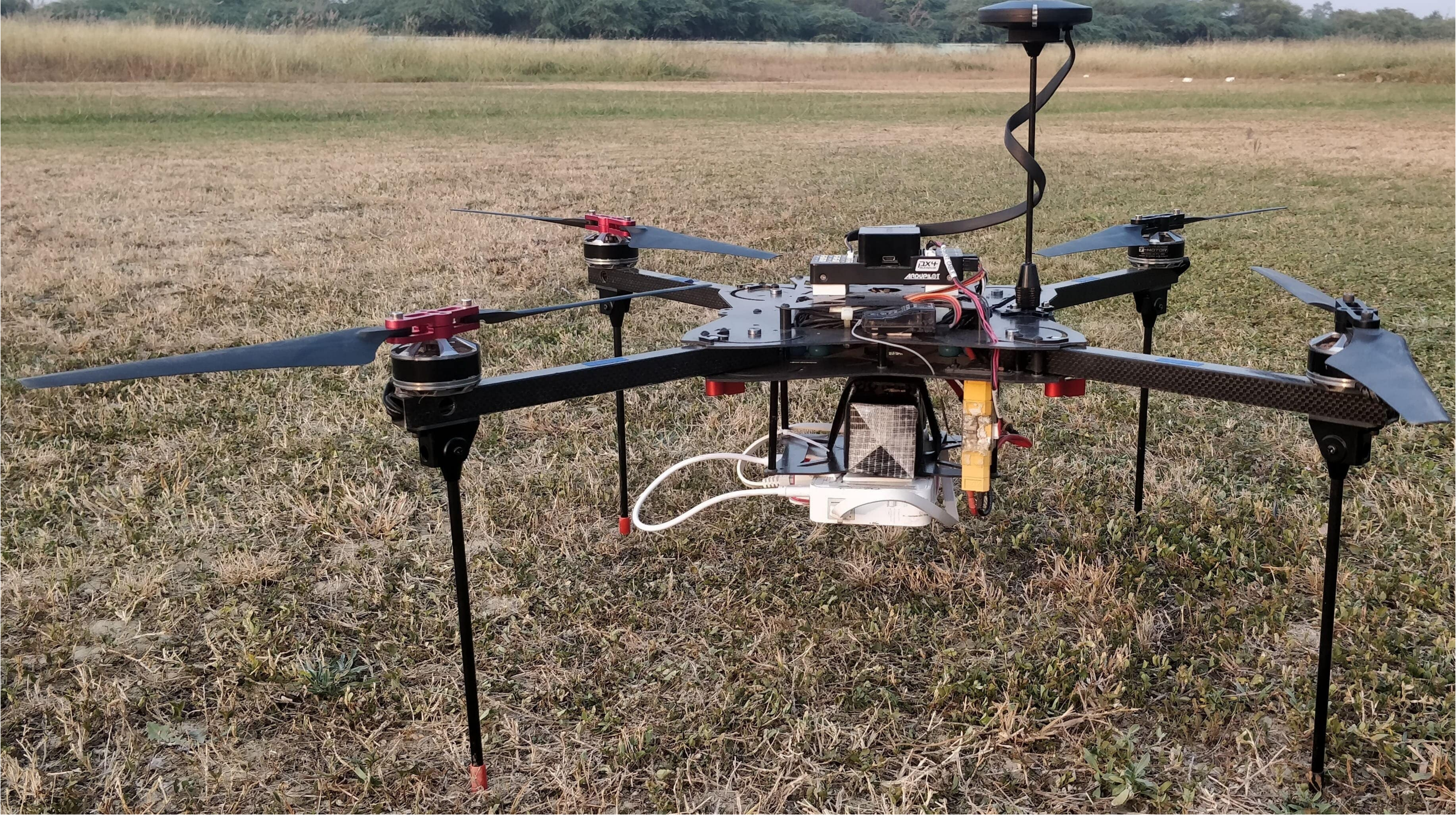}
		\label{fig_quadcopter_ground}	
	}
	\subfigure[Helicopter]
	{
		\includegraphics[width=4.2cm]{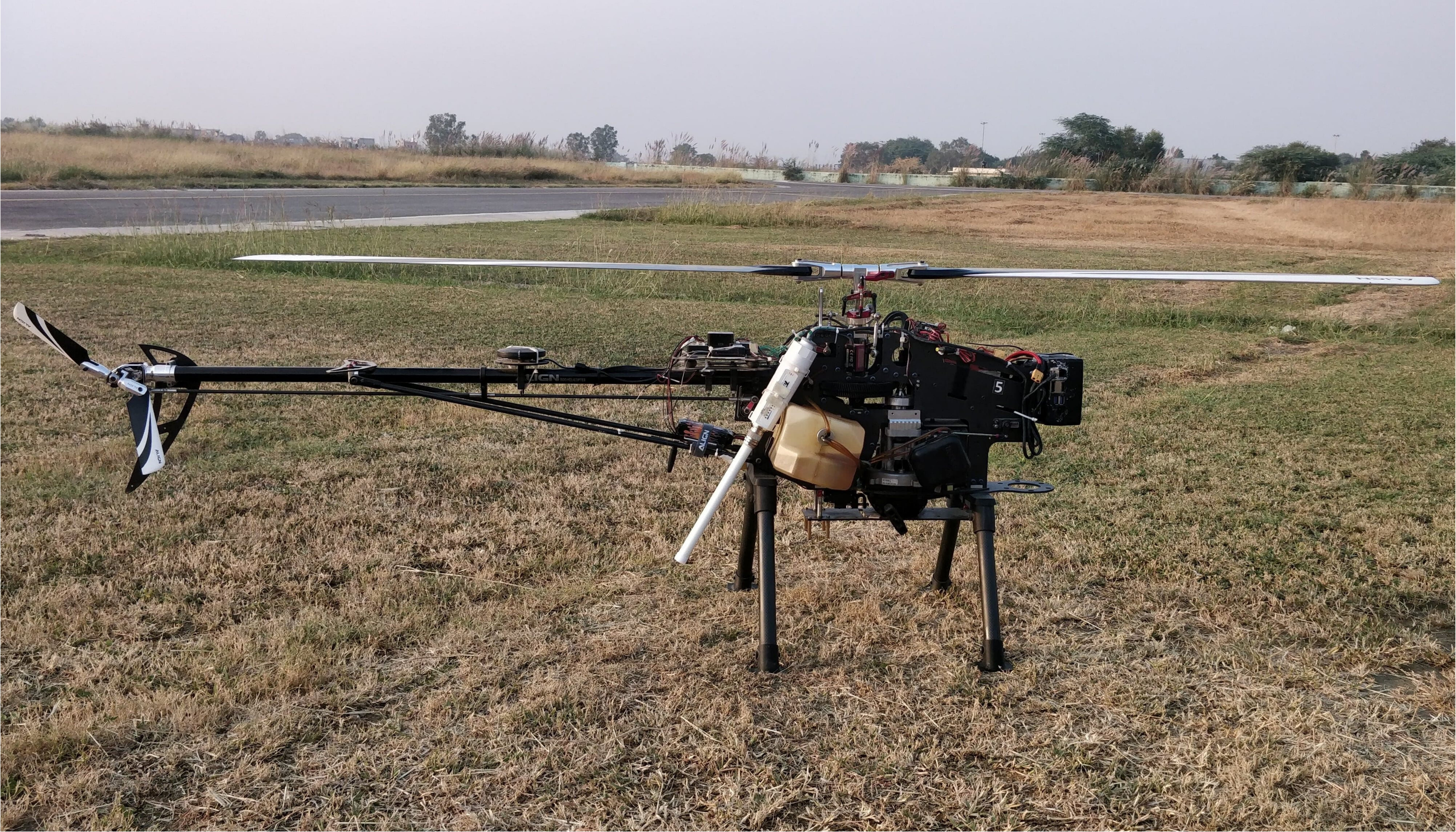}
		\label{fig_helicopter_ground}	
	}
	\caption{Implemented quadcopter and helicopter along with wireless module, autopilot, and companion computer. (a) Quadcopter. (b) Helicopter.}
	\label{fig_implemented_quard_heli_ground}
\end{figure}

\begin{figure}[!ht]
	\centering
	\hspace{-3mm}
	\subfigure[Quadcopter]
	{
		\includegraphics[width=4.2cm]{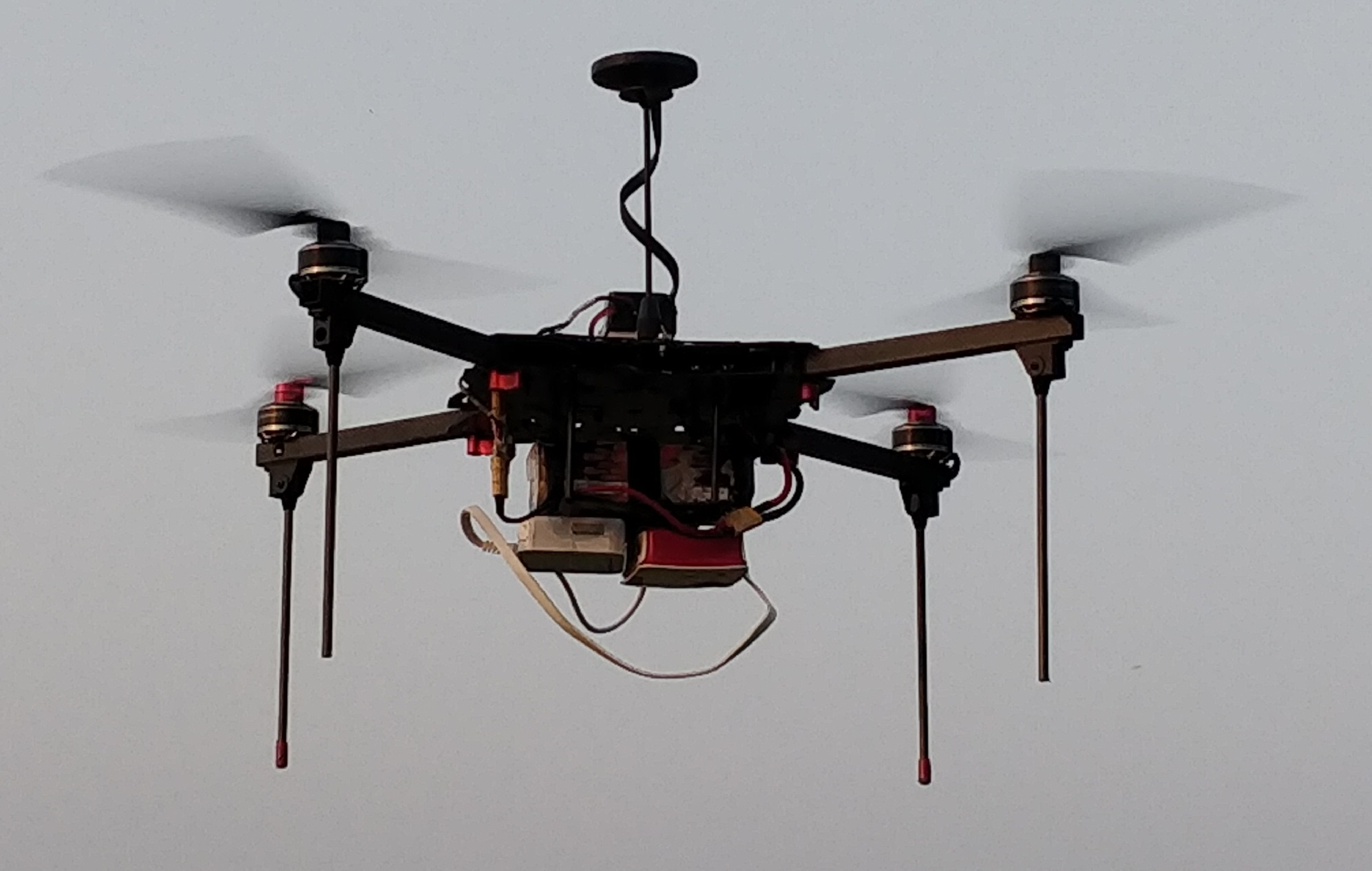}
		\label{fig_quadcopter_air}	
    }
	\subfigure[Helicopter]
	{
		\includegraphics[width=4.2cm]{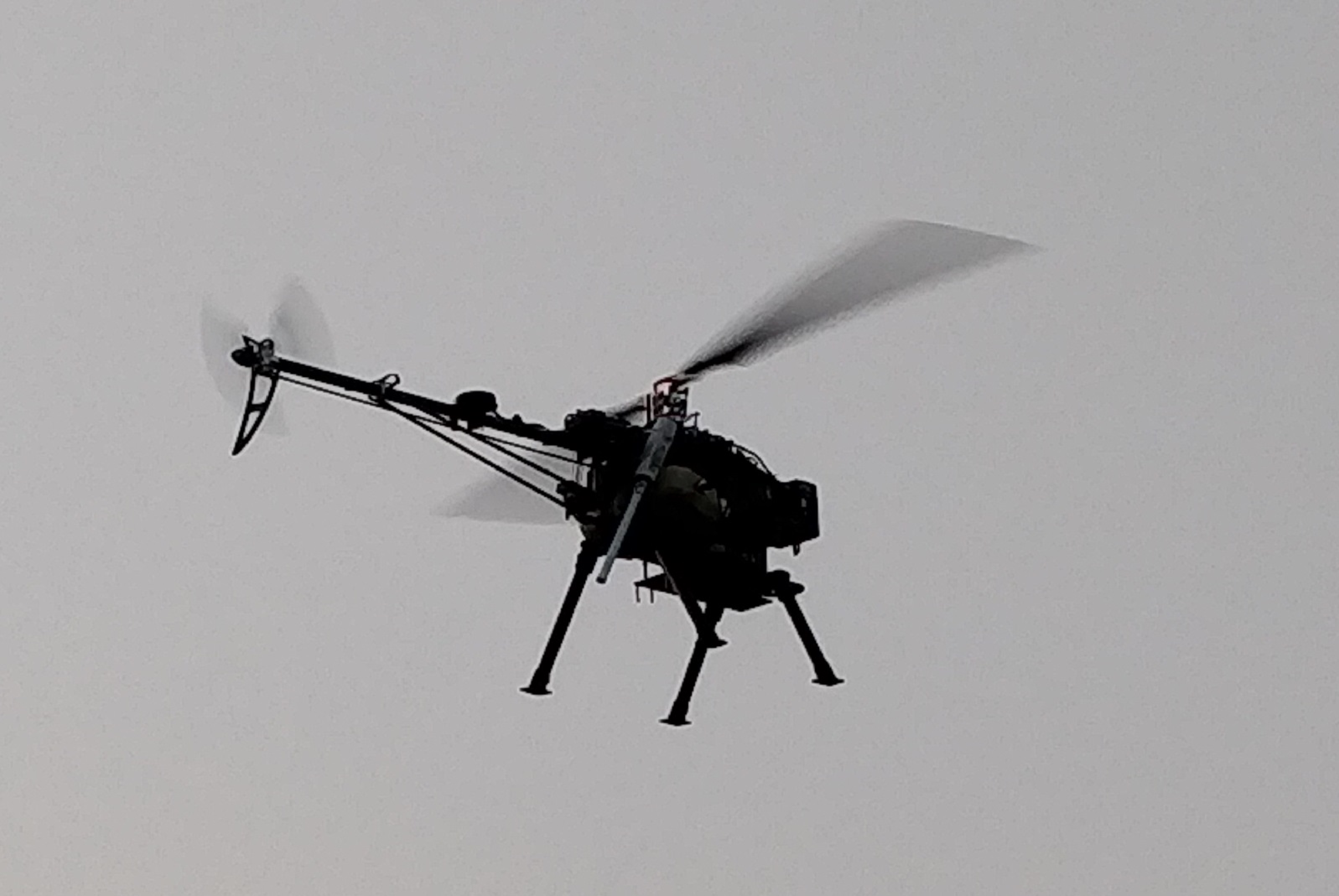}
		\label{fig_helicopter_air}	
	}
	\subfigure[Swarm of quadcopters]
	{
		\includegraphics[width=8cm]{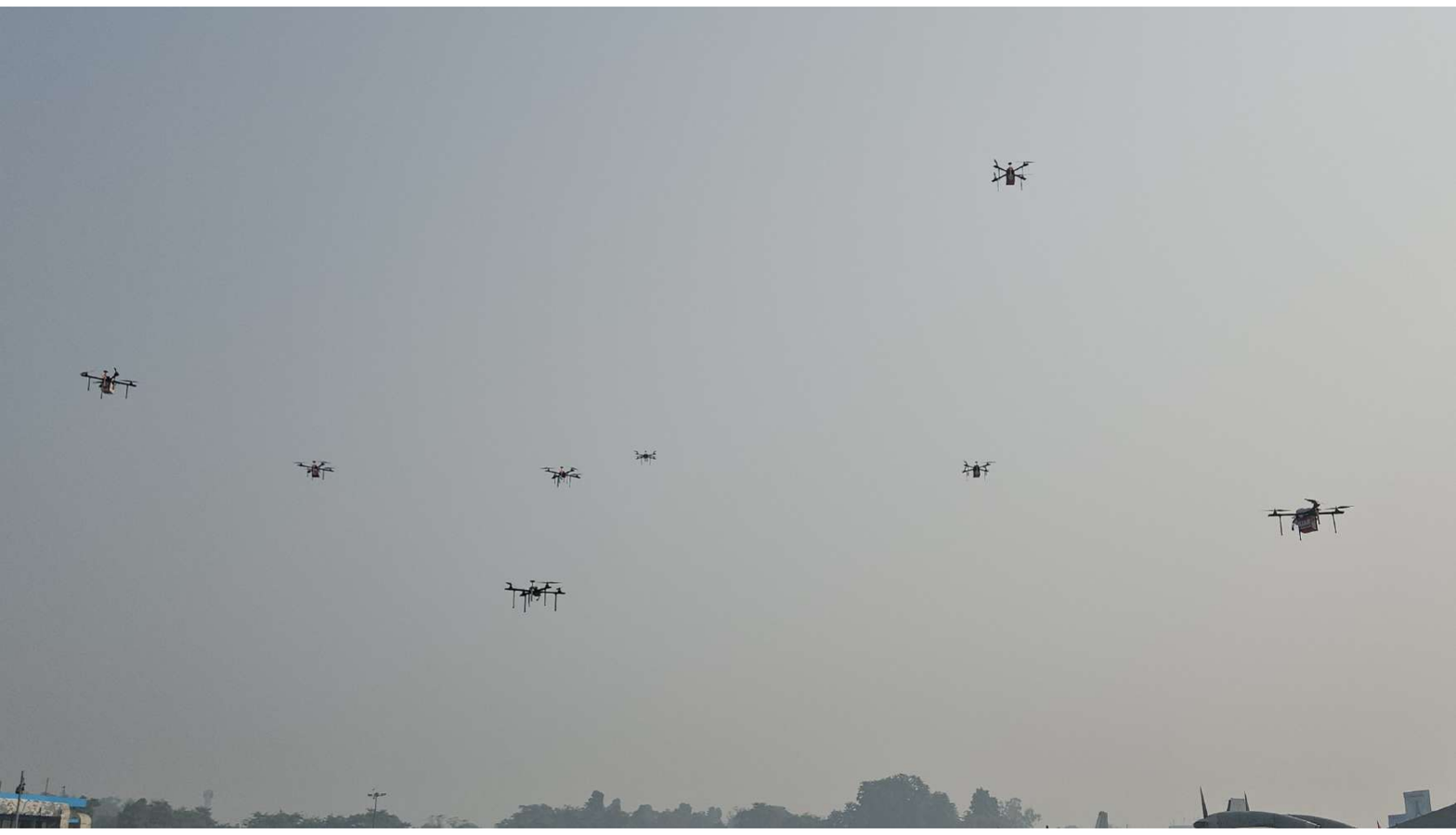}
		\label{fig_swarm}	
	}
	\caption{Deployment scenario of quadcopter and helicopter. (a) Flying a single quadcopter in the air. (b) Flying a single helicopter in the air. (c) A swarm of quadcopters is flying in the air.}
	\label{fig_implemented_quard_heli_flying}
\end{figure}

Fig. \ref{fig_quadcopter_ground} shows a quadcopter with TP-Link WR902AC Wi-Fi, Pixhawk2, and Raspberry Pi 4 Model B, preferred for short-range UAV applications. Fig. \ref{fig_helicopter_ground} depicts a helicopter with AlfaTube 2H, Pixhawk2, and $UP^2$, used for long-distance UAV gateway communication. All the modules in Table \ref{table_wireless_protocol_specifications} were tested with our UAVs. Fig. \ref{fig_implemented_quard_heli_flying} shows different deployment scenarios, including a single quadcopter (Fig. \ref{fig_quadcopter_air}) and a helicopter (Fig. \ref{fig_helicopter_air}) in flight, and swarm networking of quadcopters (Fig. \ref{fig_swarm}).

\subsection{Implementation of UAV Gateway}
\begin{figure}[t]
	\centering
	\includegraphics[width=6cm]{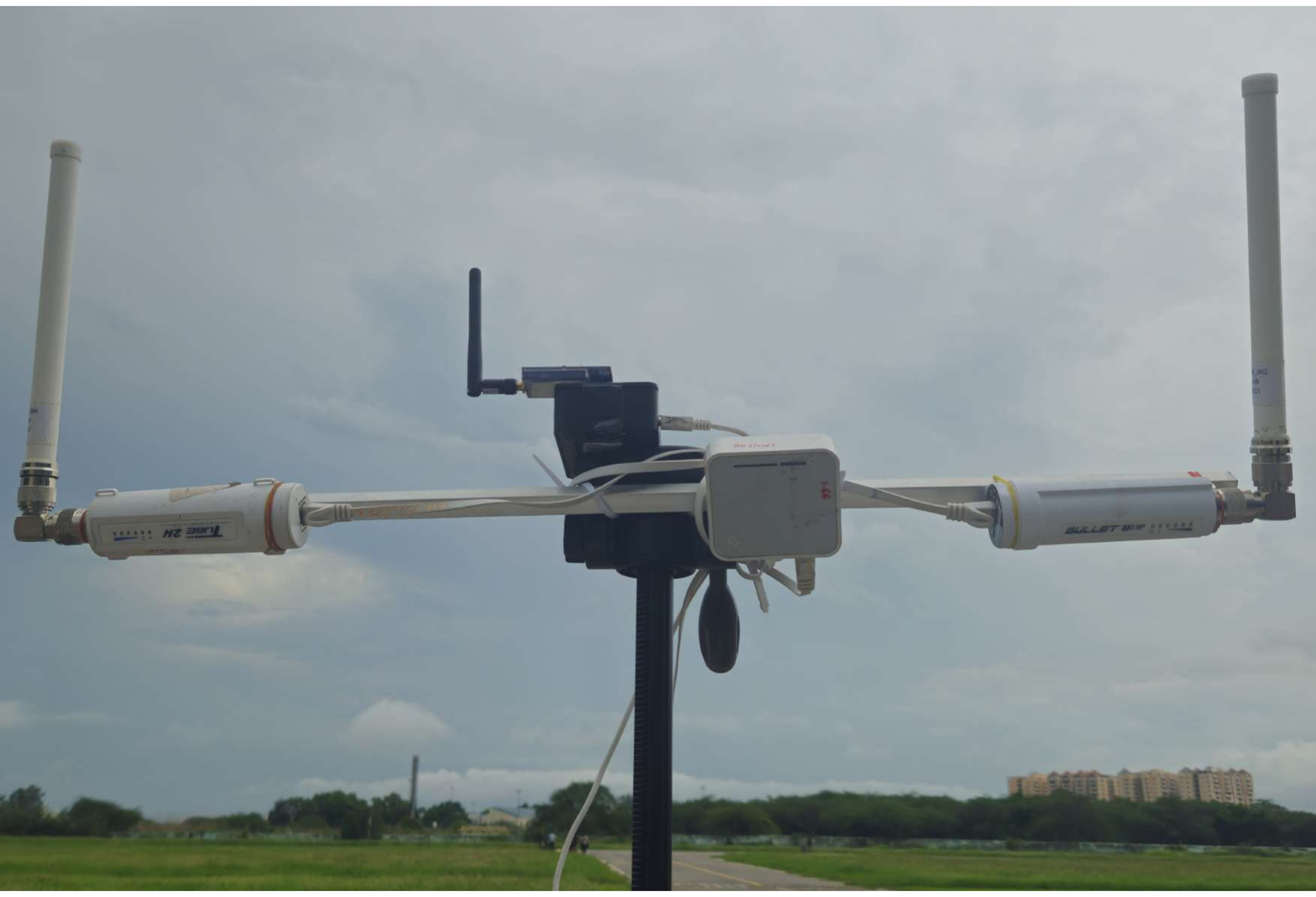}
	\caption{UAV gateway with the used components in the implementation}
	\label{fig_gateway_architecture_implementation}
\end{figure}

The components used for implementing the UAV gateway are shown in Fig. \ref{fig_gateway_architecture_implementation}. Raspberry Pi 4 Model B was used as the companion computer, running Ubuntu Server 20.04 installed on a 32 GB SD card. The Raspberry Pi was connected to four wireless modules on the UAV side -- AlfaTube 2H, Ubiquiti Bullet M2, TP-Link WR902AC, and Microhard pMDDL2450 -- via Ethernet. Users can interface with other wireless modules as required. On the other side, the Raspberry Pi was connected to a JioFi JMR540 via USB to integrate with the core Internet network. After configuration, we executed several commands to complete the UAV gateway setup: IP configuration of all modules, enabling IP forwarding, restarting the network interface, enabling IP masquerading, and saving the IP table rules, as shown in Fig. \ref{fig_gateway_configuration}.
\subsection{Implementation of Data Processing and Service Layer}
For DPSL implementation, we used ROS as middleware between the Web-based GCS and the data processing unit. Separate TCP/IP connections were created to handle incoming ROS topics from GCScom and forward incoming ROS services from users to GCScom. Python scripts were used to implement this process. The step-by-step procedures for accepting UAV connection requests and processing ROS topics and services are illustrated in Figs. \ref{fig_DPSL_server_rostopics} and \ref{fig_DPSL_server_service} using flowcharts.

\begin{figure}[t]
	\centering
	\includegraphics[width=6cm]{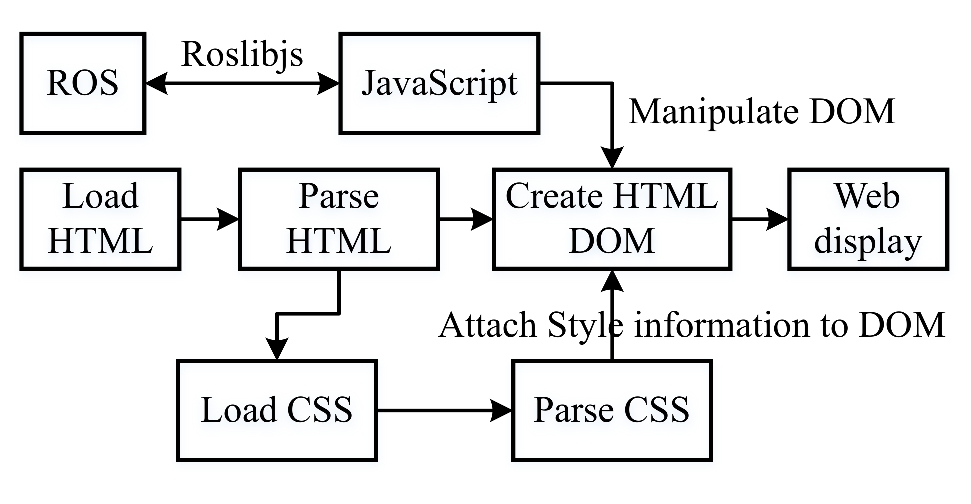}
	\caption{The functional block diagram and implementation of GCS}
	\label{fig_GCS_funcational}
\end{figure}

\begin{figure}[t]
	\centering
	\includegraphics[width=8cm]{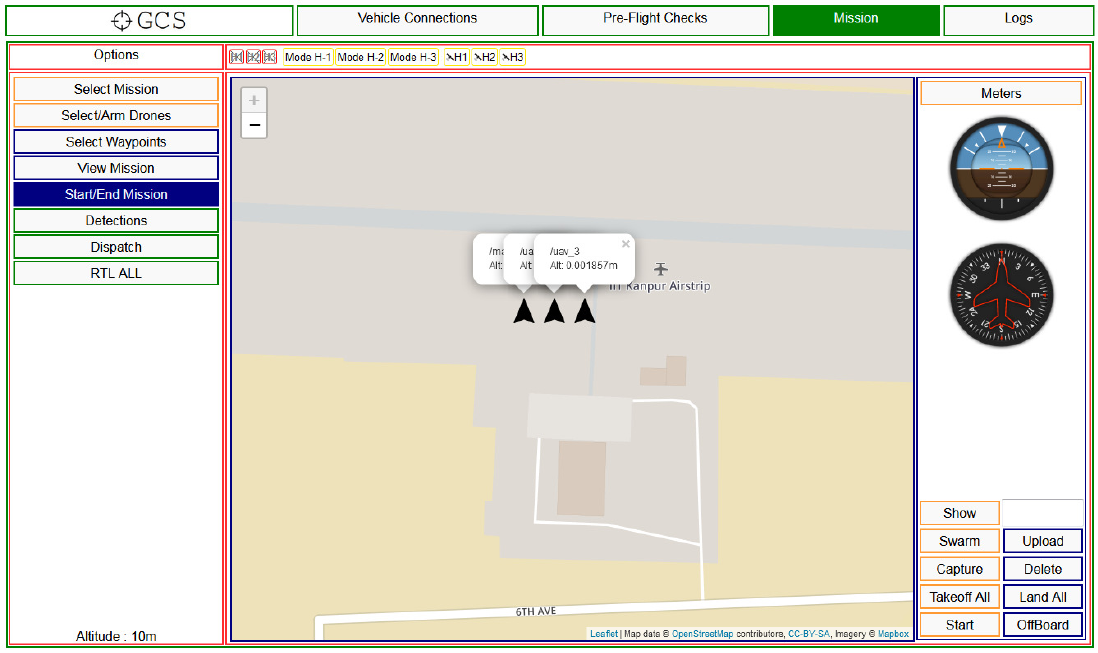}
	\caption{The Web view of GCS}
	\label{fig_GCS_webview}
\end{figure}

To design and implement the GCS, essential software tools like JavaScript, HTML, and CSS were used. Fig. \ref{fig_GCS_funcational} shows how these tools interconnect for GCS functionality. The parsed HTML and CSS create a document object model (DOM) rendered in the browser, which users view as a web page. The GCS web page interacts with ROS via JavaScript using \textit{Rosbridge}, allowing users to manipulate the DOM for interactivity. The web view of the implemented GCS is depicted in Fig. \ref{fig_GCS_webview}.

\begin{table}[]
\centering
\footnotesize
{
\caption{Adjusted/Modified Parameters in Wireless Module}
\label{table_adjustment_parameters}
\begin{tabular}{|c|ccc|}
\hline
                                              & \multicolumn{3}{c|}{Module}                                             \\ \cline{2-4} 
\multirow{-2}{*}{Parameters}                  & \multicolumn{1}{c|}{TP-Link} & \multicolumn{1}{c|}{Bullet M2} & Alfa 2H \\ \hline
Keepalive time (sec)                          & \multicolumn{1}{c|}{120}     & \multicolumn{1}{c|}{300}       & 300     \\ \hline
Keepalive probes                              & \multicolumn{1}{c|}{9}       & \multicolumn{1}{c|}{15}        & 15      \\ \hline
Probe interval (sec)                          & \multicolumn{1}{c|}{75}      & \multicolumn{1}{c|}{120}       & 120     \\ \hline
{Translation lookaside buffer (TLB)} & \multicolumn{1}{c|}{32}      & \multicolumn{1}{c|}{32}        & 16      \\ \hline
{Transmission power (dBm)} & \multicolumn{1}{c|}{22}      & \multicolumn{1}{c|}{19}        & 27      \\ \hline
{RX AMPDU time Spacing ($\mu$sec)} & \multicolumn{1}{c|}{4}      & \multicolumn{1}{c|}{8}        & 8      \\ \hline
{Short retry limit} & \multicolumn{1}{c|}{7}      & \multicolumn{1}{c|}{15}        & 15      \\ \hline
{Long retry limit} & \multicolumn{1}{c|}{4}      & \multicolumn{1}{c|}{10}        & 10      \\ \hline
{Distance setting (meter)} & \multicolumn{1}{c|}{450}      & \multicolumn{1}{c|}{2000}        & 2000      \\ \hline
\end{tabular}
}
\end{table}

\begin{table}[]
\centering
\footnotesize
\caption{Handover delay during UAV connectivity transitions between gateways using different protocols across various wireless modules was examined. The results highlight how protocol selection and ECMP routing can minimize handover delays, demonstrating the architecture's ability to maintain seamless UAV-to-gateway communication, which is crucial for reliable multi-UAV operations.}
\label{table_handover_delay}
\begin{tabular}{|c|c|c|c|}
\hline
\textbf{Protocol}     & \textbf{Module} & \textbf{G1$\rightarrow$G2} & \textbf{G2$\rightarrow$G1} \\ \hline
\multirow{3}{*}{ICMP} & TP-Link         & 0.81 $\pm$ 0.085           & 0.82 $\pm$ 0.092           \\ \cline{2-4} 
                      & Bullet M2       & 0.82 $\pm$ 0.084           & 0.81 $\pm$ 0.091           \\ \cline{2-4} 
                      & Alfa 2H         & 0.81 $\pm$ 0.084           & 0.81 $\pm$ 0.099           \\ \hline
\multirow{3}{*}{UDP}  & TP-Link         & 0.76 $\pm$ 0.078           & 0.84 $\pm$ 0.1             \\ \cline{2-4} 
                      & Bullet M2       & 0.84 $\pm$ 0.085           & 0.86 $\pm$ 0.089           \\ \cline{2-4} 
                      & Alfa 2H         & 0.82 $\pm$ 0.082           & 0.82 $\pm$ 0.086           \\ \hline
\multirow{3}{*}{TCP}  & TP-Link         & 1.00 $\pm$ 0.001           & 1.04 $\pm$ 0.082           \\ \cline{2-4} 
                      & Bullet M2       & 0.99 $\pm$ 0.005           & 0.99 $\pm$ 0.008           \\ \cline{2-4} 
                      & Alfa 2H         & 1.00 $\pm$ 0.003           & 0.99 $\pm$ 0.003           \\ \hline
\end{tabular}
\end{table}
\section{Performance Evaluation}\label{sec_performance_evalution}
\subsection{Experimental Setup}\label{ssec_experimental_setup}
We conducted experiments with UNet at the airstrip in the flight laboratory at IIT Kanpur from December 2023 to July 2024. The performance of UNet was assessed under various conditions, including in-flight and ground data collection modes. The experiments used five quadcopters (UAVs), two gateways, and one GCS. In specific experiments, the UAVs operated using a mesh topology as a swarm. In contrast, in others, each UAV was individually directly connected to a gateway. In the swarm formation experiments, the UAVs flew in a V-shaped formation with 20-meter inter-UAV spacing and a 120-degree formation angle, with each UAV maintaining a $\pm$3-meter altitude difference from its neighbors for collision avoidance, with the formation advancing up to 1 km from the gateway. For gateway performance evaluation, a single UAV was employed to isolate individual metric performance, moving within a 300-meter radius of the gateway throughout the experiment. The GCS was connected to the gateways either via Ethernet (LAN) or JioFi JMR540 4G modem. In the experiment, we used three wireless modules to communicate between UAVs and gateway: TP-link WR902AC, Ubiquiti Bullet M2, and AlfaTube 2H. We used 95\% confidence intervals as the error bars for selected performance metrics in Table~\ref{table_handover_delay} and Figs.~\ref{plot_gw_performance}, \ref{plot_e2e_performance}, \ref{plot_task_execution_time}, and \ref{plot_current_consumption_bullet}. Table \ref{table_adjustment_parameters} shows the adjustment of various parameters for each wireless module in its configuration, helping them perform better in different performance metrics. The impact of these adjustments will be analyzed in the results section. The `distance setting' decides whether the module communicates for a short or long distance based on its value, which proportionally sets acknowledgment time and other factors in the MAC layer. Note that `distance setting' does not imply the exact distance of communication.

Note, we evaluate key performance metrics (e.g., handover delay, end-to-end delay, packet loss probability, throughput, protocol behavior, traffic requirements, task execution time, and reliability) that inherently capture the requirements of diverse applications, instead of evaluating individual applications.

The source code of data collection and performance analysis used in the paper is available in \url{https://github.com/samshadnc/UNET}.

\subsection{Results and Discussion}\label{ssec_results_discussion}
\subsubsection{Handoff Delay}
Table \ref{table_handover_delay} presents the handover delay experienced when a UAV transitions its connectivity from one gateway to another during flight. This delay was analyzed from the perspective of the UAV as it moved between different locations. Our experiments evaluated three communication protocols -- TCP, UDP, and ICMP -- across various wireless modules. Each protocol was selected for its distinct data transfer strategies, reflecting their use in UAV-to-gateway communications. For this experiment, we employed two gateways, designated G1 and G2. The notation G1$\rightarrow$ G2 denotes the handover of a UAV's connection from G1 to G2, while G2$\rightarrow$ G1 signifies the handover from G2 to G1. Equal-cost multi-path routing (ECMP) was used to facilitate better handover of the existing connection between gateways.

Table \ref{table_handover_delay} illustrates that TCP experiences slightly higher handover delays than UDP and ICMP across all wireless modules. The performance is because TCP maintains a secure virtual connection during data transfer and requires re-establishing the connection before starting new data transfers, unlike UDP and ICMP. From the wireless modules' perspective, performance is generally consistent across all protocols, indicating similar efficiencies due to using the same wireless standard. Overall, UNet consistently achieves minimal handover delays across all protocols in highly mobile and dynamic environments. This performance is attributed to its effective identification of existing and non-existing paths using ECMP, as well as the optimization of other network parameters like retransmission time.
\begin{figure}[h!]
    \centering
    \subfigure[Throughput]{
        \includegraphics[width=6cm]{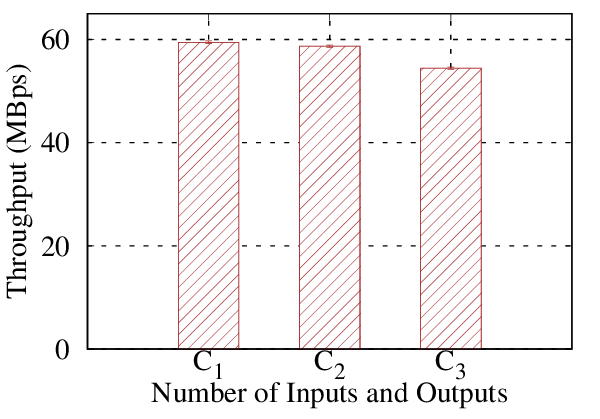}
        \label{plot_gw_throughput}
    }\quad\hspace{-0.8cm}
    \subfigure[Data processing delay]{
        \includegraphics[width=6cm]{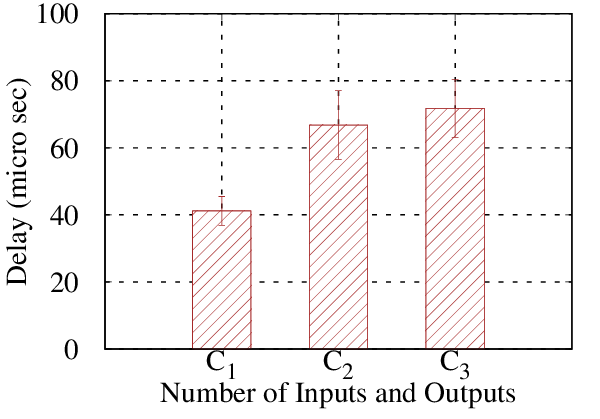}
	   \label{plot_gw_delay}
    }
    \caption{Gateway performance under three different input/output configurations. The results demonstrate the gateway's optimal performance with higher throughput and lower processing delay in the one-input scenario, highlighting its efficiency and scalability in handling multiple data streams while maintaining minimal delay, crucial for the reliable operation of the UNet architecture.}
    \label{plot_gw_performance}
\end{figure}
\begin{figure*}[ht!]
	\centering
   \subfigure[]{
		\includegraphics[width=6cm]{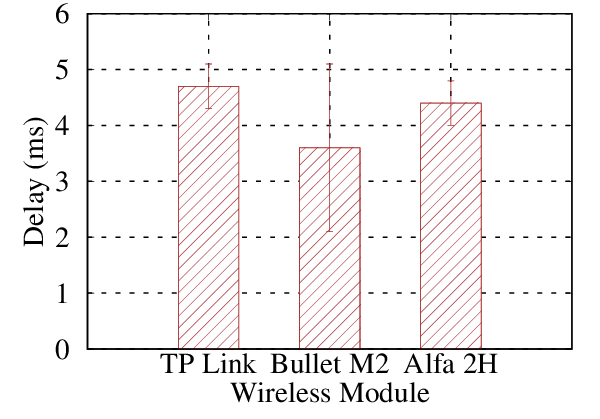}
		\label{plot_e2e_delay_tcp}
	}
	\quad\hspace{-0.9cm}
	\subfigure[]{
		\includegraphics[width=6cm]{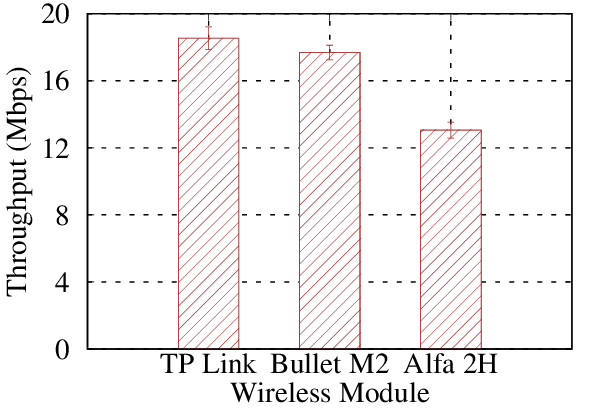}
		\label{plot_e2e_throughput_tcp}
	}
        \quad\hspace{-0.9cm}
	\subfigure[]{
		\includegraphics[width=6cm]{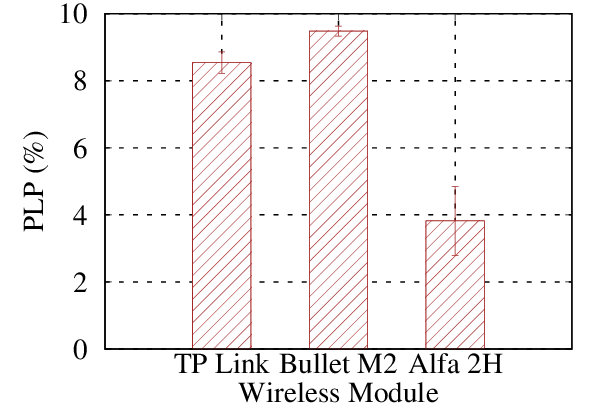}
		\label{plot_e2e_plp_tcp}
	}
	\caption{End-to-end performance for (a) delay, (b) throughput, and (c) packet loss probability. UNet achieves low packet delivery delay, high throughput, and minimal packet loss probability, demonstrating its reliability and efficiency. AlfaTube 2H prioritizes reliability for long-distance communication, Ubiquiti Bullet M2 balances throughput and reliability, and TP-Link WR902AC excels in short-range scenarios with higher throughput. These results highlight UNet's adaptability to diverse UAV communication environments.}
	\label{plot_e2e_performance}
\end{figure*}
\subsubsection{Gateway Performance}
The gateway's performance is crucial to UNet. We evaluate key performance metrics to ensure optimal functionality, including throughput and data processing delay. As mentioned earlier, the gateway has multiple inputs and one output. From the gateway's perspective, throughput is defined as the volume of input data successfully transferred to the output port for delivery to the GCS. Data processing delay refers to the time required to process input data and transfer it from the input port to the output port.

In this experimental setup, we evaluated the gateway's performance under three scenarios: (C$_1$) one input and one output, (C$_2$) two inputs and one output, where only one input transmits data at a time, and (C$_3$) two inputs and one output, with both inputs transmitting data simultaneously.

Figs. \ref{plot_gw_throughput} and \ref{plot_gw_delay} illustrate the gateway's throughput and data processing delay, respectively. As expected, the gateway shows significantly higher throughput and lower processing delay in the single input/output scenario compared to the other cases. Conversely, the gateway requires slightly more processing time and exhibits reduced throughput in the two-input/single-output scenario due to the need to process data from both inputs before transferring it to the output. Overall, the gateway maintains high throughput and minimal data processing delay across all scenarios, underscoring the superiority of its performance.

\subsubsection{End-to-end (UAV to GCS)}
Evaluating the end-to-end performance between UAVs and GCS is crucial for understanding the reliability and efficiency of the communication system. This evaluation covers various network metrics, including packet delivery delay (PDD), throughput, and packet loss probability (PLP), as illustrated in Fig. \ref{plot_e2e_performance}. This study tested different wireless modules between the gateway and the UAV. At the same time, a LAN connection was used between the gateway and the GCS to ensure stability. Data collection for this experiment was conducted using TCP, known for its reliability and ordered data delivery.

UNet demonstrated significantly low PDD (Fig. \ref{plot_e2e_delay_tcp}), indicating rapid packet delivery from the UAV to the GCS. It achieved high throughput (Fig. \ref{plot_e2e_throughput_tcp}), reflecting efficient transmission of large volumes of data over the network, and exhibited low PLP (Fig. \ref{plot_e2e_plp_tcp}), suggesting minimal packet loss during transmission. These metrics collectively indicate that UNet provides a highly reliable network, even in the dynamic and mobile environment typical for UAV operations.

An interesting observation is that AlfaTube 2H's throughput, shown in Fig. \ref{plot_e2e_throughput_tcp}, is the lowest among the tested modules. However, AlfaTube 2H's PLP, depicted in Fig. \ref{plot_e2e_plp_tcp}, is also the lowest. Also, AlfaTube 2H experiences higher packet delivery delays than the other modules, as illustrated in Fig. \ref{plot_e2e_delay_tcp}. Despite using the standard Wi-Fi protocol, all three modules' performance varies significantly due to design optimizations.

AlfaTube 2H is optimized for long-distance communication. It adjusts several parameters in the Medium Access Control and Logical Link Control sublayers of the data link layer: a larger contention window size to reduce collisions, frequent use of the Request to Send/Clear to Send mechanism to avoid hidden node issues, enhanced flow control mechanisms for retransmissions and error correction, and optimized frame aggregation for efficient long-distance communication. AlfaTube 2H has a slightly higher retransmission timeout, which reduces throughput but increases the packet delivery success rate. Thus, AlfaTube 2H transmits less data with a high success rate, resulting in a lower PLP, as shown in Figs. \ref{plot_e2e_throughput_tcp} and \ref{plot_e2e_plp_tcp}.

Although Ubiquiti Bullet M2 is also designed for long-distance communication, it is less restrictive with retransmissions than AlfaTube 2H. As a result, Ubiquiti Bullet M2 sends more packets but with a lower success rate. TP-Link WR902AC, designed for short-range and indoor use, achieves higher throughput than AlfaTube 2H and Ubiquiti Bullet M2, reflecting its optimization for environments where packet loss and delay are less critical. These findings highlight the importance of tailoring communication modules to their operational environments, balancing throughput, reliability, and packet delivery delay to achieve optimal performance.

\subsubsection{Task Execution time}
\begin{figure}[h!]
	\centering
	\includegraphics[width=6cm]{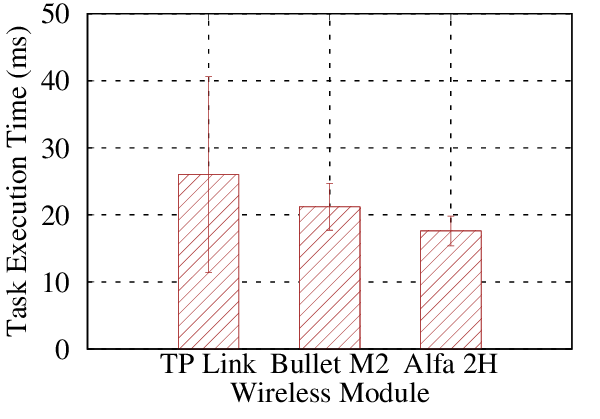}
	\caption{Task execution time. The architecture minimizes execution time for UAV commands initiated from the GCS, with AlfaTube 2H and Ubiquiti Bullet M2 outperforming TP-Link WR902AC due to superior external antennas. These results demonstrate the architecture's efficiency in reducing delays ensuring responsive UAV operations.}
	\label{plot_task_execution_time}
\end{figure}

Fig. \ref{plot_task_execution_time} shows the task execution delay experienced by the UAV after a command was initiated from the GCS. This delay includes the total time required for the task to be executed on the UAV, starting from when the command was sent. It consists of the transmission time from the GCS to the gateway, then to the UAV, and the time for the UAV's onboard computer to receive an acknowledgment from the flight controller after successfully completing the task. The transmission time, which varies depending on the wireless module used, is a key factor in this delay.

For testing, we used two commands: arm throttle and flight mode change. The figure demonstrates that our architecture significantly reduces task execution time on the UAV. However, it also shows that the TP-Link WR902AC experiences a slightly higher delay than the AlfaTube 2H and Ubiquiti Bullet M2. This difference is likely due to the type of antenna used in these modules. The TP-Link relies on an onboard PCB antenna, which typically provides weaker signal strength and is more susceptible to interference. In contrast, the AlfaTube 2H and Ubiquiti Bullet M2 use external antennas that offer superior signal quality and coverage, leading to faster data transmission. The onboard antenna in the TP-Link module likely accounts for the observed increase in delay.
\begin{figure}[h!]
	\centering
	\includegraphics[width=7cm]{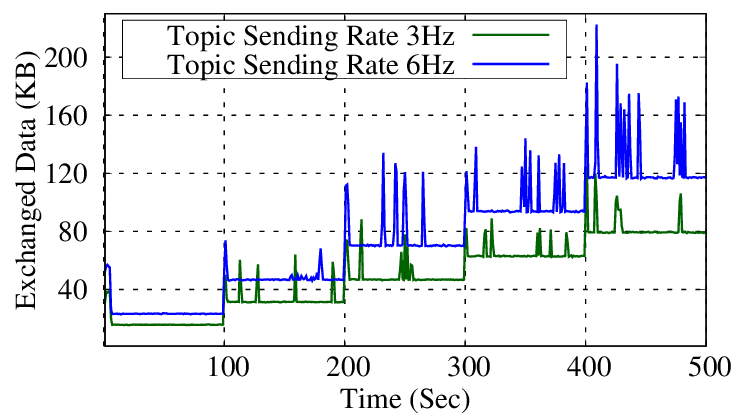}
	\caption{Exchanged traffic over time. Traffic is exchanged between a gateway and UAVs over time as new UAVs join every 100 seconds in a mesh network. The architecture efficiently handles increasing traffic, with spikes reflecting the addition of UAVs and mesh connectivity maintenance. Results highlight scalability and robust management of multi-UAV traffic in dynamic networks.}
	\label{plot_exchangeTraffic}
\end{figure}
\begin{figure*}[t]
	\centering
    \subfigure[]{
		\includegraphics[width=6cm]{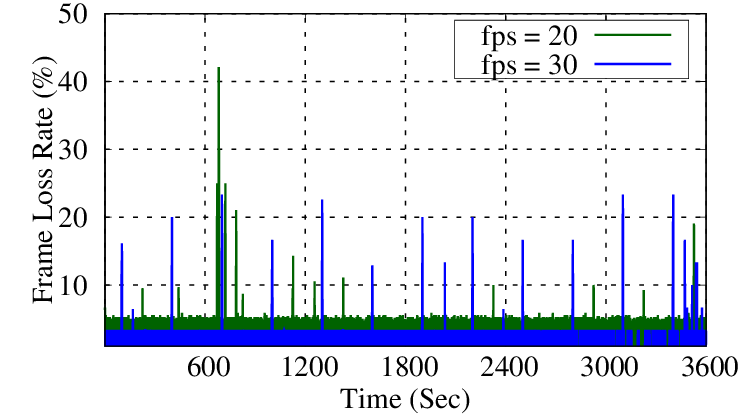}
		\label{plot_reliability_framerate}
	}
	\quad\hspace{-0.9cm}
   \subfigure[]{
		\includegraphics[width=6cm]{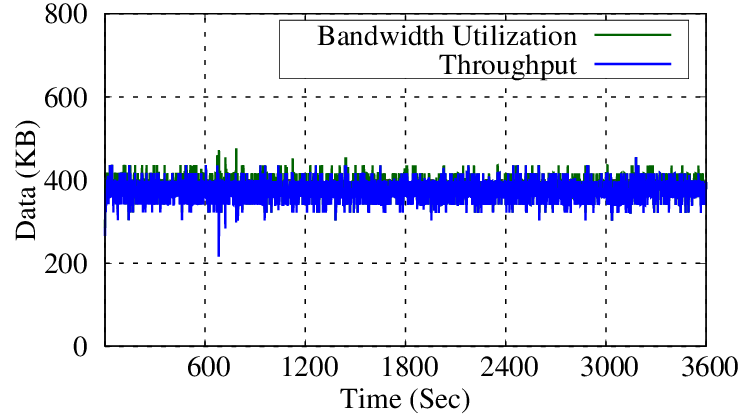}
		\label{plot_reliability_throughputUtilization20fps}
	}
	\quad\hspace{-0.9cm}
	\subfigure[]{
		\includegraphics[width=6cm]{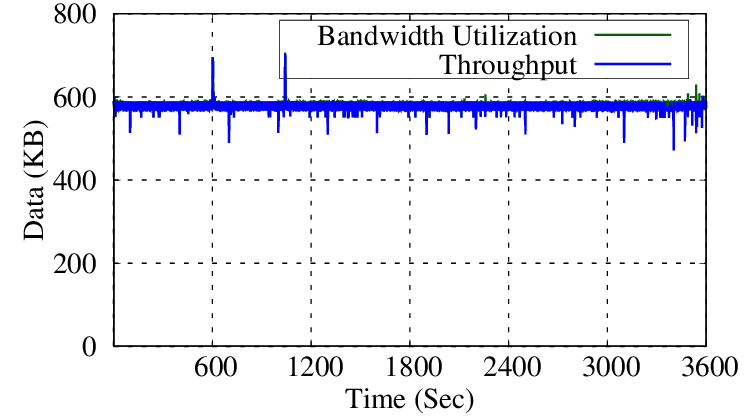}
		\label{plot_reliability_throughputUtilization30fps}
	}
	\caption{Reliability performance. (a) frame loss rate. (b) throughput vs bandwidth utilization for 20 fps. (c) throughput vs bandwidth utilization for 30 fps. Results show minimal frame rate loss and near-identical throughput and bandwidth utilization for 20 fps and 30 fps video streams, highlighting UNet's efficiency and robustness in maintaining consistent and reliable data transmission.}
	\label{plot_reliability}
\end{figure*}
\subsubsection{Exchanged traffic over time}
Fig. \ref{plot_exchangeTraffic} shows the traffic exchanged between a gateway and multiple UAVs over time, with a new UAV added every 100 seconds. Initially, for the first 100 seconds, only one UAV is connected to the gateway. At the 100-second mark, an additional UAV is introduced, resulting in two UAVs transmitting data to the gateway between 100 and 199 seconds. This pattern continues in subsequent intervals, adding UAV every 100 seconds. This experiment aims to demonstrate how the architecture manages the connectivity and traffic of multiple UAVs under a single gateway.

The UAVs and gateway are connected in a mesh network in this setup. Each UAV transmits two ROS topics: global position and orientation and the battery's remaining voltage. We tested two different sending frequencies for each topic: 3 Hz and 6 Hz.

The figure demonstrates that the architecture efficiently handles the data traffic generated by multiple UAVs in a mesh topology. At each 100-second interval, there is a noticeable spike in traffic, signaling the addition of a new UAV to the network. These spikes indicate the extra traffic required to establish a mesh connection with the gateway within the multiple master ROS framework, in addition to regular topic sharing.

Additional spikes in exchanged traffic can also be observed, attributed to the ongoing maintenance of mesh connectivity within the network. The figure further reveals an interesting pattern: the magnitude of these spikes increases as more UAVs join the network. This increase is due to the additional overhead of data transmission among the UAVs to maintain mesh connectivity and update their routing tables.

Initially, at the 100-second mark, there is only one UAV, and the data traffic remains relatively stable except for the initial spike when the UAV joins the gateway. As more UAVs join at subsequent 100-second intervals, the magnitude of the spikes increases, reflecting the extra traffic contributed by each new UAV to maintain the mesh network and update routing information.

\subsubsection{Reliability}
To assess the reliability of UNet, we conducted an experiment involving continuous video data transmission from a UAV to a GCS. Building upon previous performance metrics, we measured reliability using a combination of frame rate loss, throughput, and bandwidth utilization, as shown in Fig. \ref{plot_reliability}.

Frame rate loss is defined as the percentage of frames lost per second. Bandwidth utilization refers to the total data transmitted over a given period, while throughput denotes the total data received within the same time frame. For this experiment, we used a video resolution of 320x240 pixels and tested two different frame rates: 20 frames per second (fps) and 30 fps.

Fig. \ref{plot_reliability_framerate} demonstrates that a nearly consistent frame rate with minimal loss indicates superior network performance for both frame rates. On average, the frame loss per second was 0.57 at 20 fps and 0.46 at 30 fps. The average frame loss rate in percentage was 2.81\% and 1.48\% at 20 fps and 30 fps, respectively. The higher loss rate at 20 fps can be attributed to the lower total frame count per second, which means each lost frame has a proportionally greater impact on the overall loss rate than at 30 fps. Figs. \ref{plot_reliability_throughputUtilization20fps} and \ref{plot_reliability_throughputUtilization30fps} depict throughput versus bandwidth utilization for 20 fps and 30 fps, respectively. These figures show that bandwidth utilization and throughput are nearly identical, indicating minimal data loss and high efficiency.

Overall, the results presented in these figures demonstrate UNet's high reliability. The nearly identical bandwidth utilization and throughput values, combined with minimal frame rate loss, underscore the architecture's efficiency and robustness in maintaining consistent video data transmission.

\begin{figure}[h!]
	\centering
	\includegraphics[width=6cm]{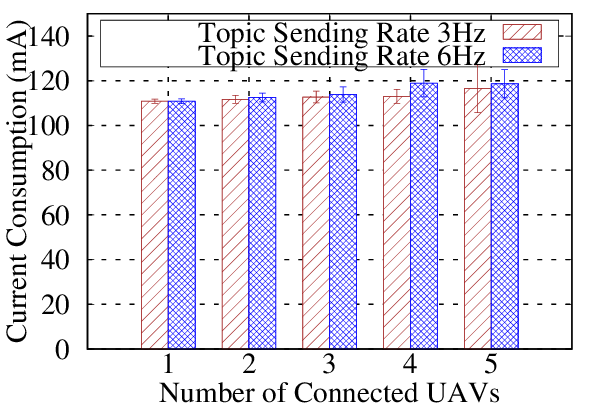}
	\caption{Current consumption of Ubiquiti Bullet M2}
	\label{plot_current_consumption_bullet}
\end{figure}
\subsubsection{Energy Consumption}
In the quadcopter platform used in this experiment, energy consumption can be decomposed into three components. The base quadcopter operation, encompassing the motors, electronic speed controllers (ESCs), and flight controller, accounts for the dominant share, approximately 79\% of total power draw (approx. 448 W), as the majority of energy is expended overcoming gravity and maintaining stable flight through continuous rotor thrust. When a communications module is integrated into the UAV, two additional energy costs arise: the mechanical cost of carrying the extra mass, and the electrical cost of powering the module itself. The communications module used in this work, comprising the RF modem, antenna, and onboard computer, weighs 500 g in total, which forces the motors to draw additional power to sustain lift, introducing a motor load penalty of approximately 20\% of total power (approx. 115 W). In contrast, the actual electrical draw of the wireless communication subsystem is negligible, contributing less than 1\% of total power (approx. 5 W), confirming that the dominant energy cost of integrating a communications module into a UAV stems from its physical mass rather than its power consumption. The flight platform was powered by a 6S lithium-ion battery (22.2 V, 8000 mAh), yielding an observed average flight time of 19 minutes during formation flight at 3 m/s under nominal conditions. Flight time is inherently subject to variability due to factors such as wind conditions, altitude, flight speed, and quadcopter maneuver patterns; however, the primary determinant remains battery energy density and total system weight. Flight endurance can be improved through higher energy-density cells, lighter payload components, and optimized flight conditions. In contrast, the helicopter variant employed in this system uses a 15cc gasoline engine, which affords significantly greater payload capacity and endurance, making it better suited for missions requiring extended operation or heavier sensor payloads. To give an overview of the energy consumption of the wireless module used in the experiment, the current consumption of the Ubiquiti Bullet M2 is discussed below.

Fig. \ref{plot_current_consumption_bullet} illustrates the current consumption of the Bullet M2 wireless module as the number of transmitting UAVs varies. We tested two different ROS topic transmission frequencies: 3 Hz (approximately 15 kBps) and 6 Hz (approximately 30 kBps). In our experimental setup, the battery voltage was maintained at 12 V. The figure shows that transmitting ROS topics results in minimal additional current consumption. This is because, on average, the wireless module consumes 108.964 mA in normal mode to maintain the mesh topology, even without any ROS topic transmission from the companion computer. Additionally, as previously discussed, the module supports long-distance communication. It is important to note that energy or current consumption is not the primary focus of the proposed architecture in this paper.

\section{Conclusion}\label{sec_conclusion_future_work} 
We presented the design of a novel, generic multi-UAV communication and networking system architecture that concurrently supports heterogeneous applications. UNet demonstrates seamless handling of infrastructure-based and infrastructure-less networks within the same framework. Additionally, we introduced a UAV gateway capable of interoperating with various wireless protocols while offering a remote monitoring and control system that enhances user convenience in managing individual applications. The practical implementation and evaluation results underscore the effectiveness of UNet.

From our design and implementation, the key takeaway is that combining ad hoc and infrastructure networks within a unified architecture, along with middleware-based data handling and network parameter optimization, is essential for supporting heterogeneous UAV applications. In addition, the use of standard interfaces and protocols enables platform-agnostic deployment and improves scalability across different UAV systems.

This work enhances the efficiency and reliability of multi-UAV systems, critical for disaster management, precision agriculture, infrastructure monitoring, etc. It optimizes task execution and data transmission, enabling real-time operations, scalability, and faster decision-making in collaborative UAV scenarios.

The next step is to develop mechanisms to support more UAVs while maintaining performance and reliability. Apart from that, another possible extension could be on developing a formal analytical model of UNet to enable theoretical analysis, including performance bounds and scalability evaluation under different network conditions. Another possible direction is evaluating network behavior under battery depletion and UAV failure conditions to further enhance the assessment of UNet resilience and adaptability. Additionally, implementing AI-driven algorithms for dynamic network optimization, predictive maintenance, and autonomous decision-making could further enhance the architecture's adaptability and efficiency. These advancements will ensure that the system can effectively handle increasing complexity and operational demands.
\section*{Acknowledgement}
We thank Ronak Jain for designing the Web-based GCS, submitted as an MTech thesis at IIT Kanpur, India.
\bibliographystyle{IEEEtran}
\bibliography{UNet}
\begin{IEEEbiography}[{\includegraphics[width=1in,height=1.25in,clip,keepaspectratio]{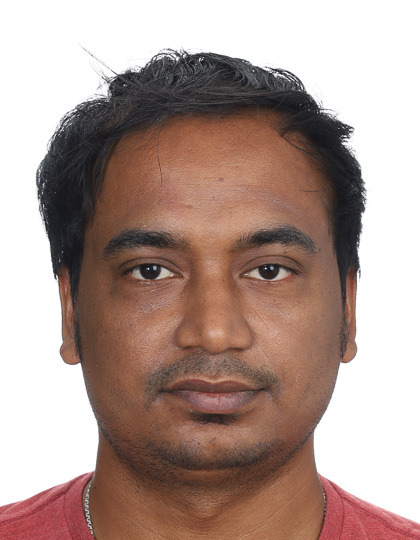}}]{Sanku Kumar Roy} (S'15) received his B.Tech. from the Maulana Abul Kalam Azad University of Technology, India, in 2012, and his MS (by Research) from the Indian Institute of Technology (IIT) Kharagpur, India, in 2019. He is currently pursuing a Ph.D. in the Department of Computing Science at the University of Alberta.

He has industry and research experience, serving as a Product Development and Support Engineer at Pervcom Consulting Pvt. Ltd. (2012--2013), a Project Assistant at IIEST Shibpur (2013--2014), and a Junior Research Fellow at IIT Kharagpur (2014--2017) and IIT Kanpur (2018--2020). His current research interests include Encrypted Traffic Analysis, Embedded AI, Low-Complexity Data Compression, Network Optimization, Internet of Things, and Wireless Sensor Networking.
\end{IEEEbiography}

\begin{IEEEbiography}[{\includegraphics[width=1in,height=1.25in,clip,keepaspectratio]{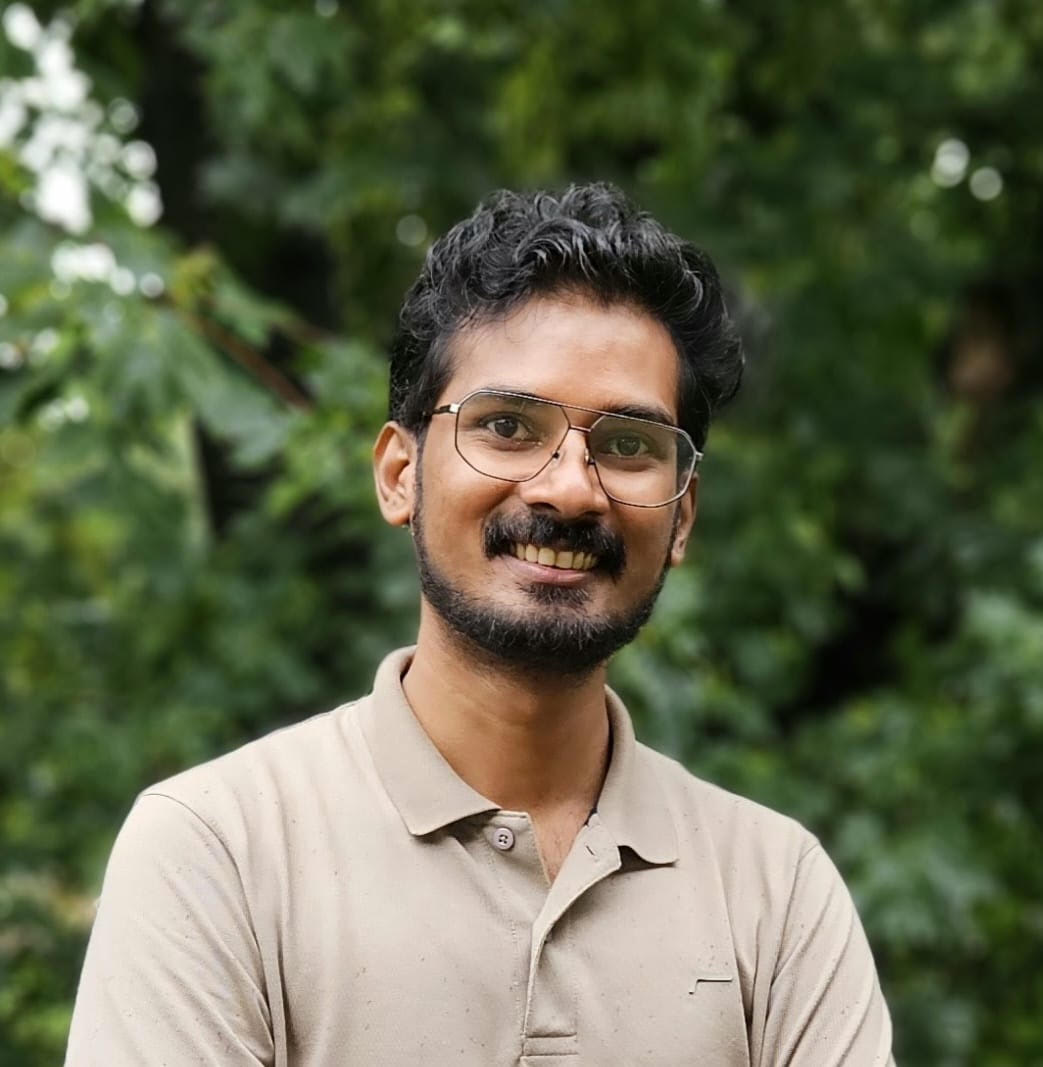}}]{Mohamed~Samshad} received his B.Tech. degree in Electronics and Communication Engineering from the University of Kerala, India, and his MS (by Research) degree in Electrical Engineering from the Indian Institute of Technology (IIT) Kanpur, India.

He has industry and research experience as an Electronic Engineer at Kasperob Robotics (2017--2019) and a Research Assistant at ICFOSS (2019--2021), and led the Multi-UAV Research Group at the SPIN Laboratory, IIT Kanpur, as a UAV Network Engineer (2021--2025). He is currently a Research Engineer at the Advanced Control Technology (ACT) Laboratory, Department of Electrical and Computer Engineering, National University of Singapore. His current research interests include computer vision, visual SLAM, 3D scene graphs, multi-robot systems, and autonomous systems.
\end{IEEEbiography}

\begin{IEEEbiography}[{\includegraphics[width=1in,height=1.25in,clip,keepaspectratio]{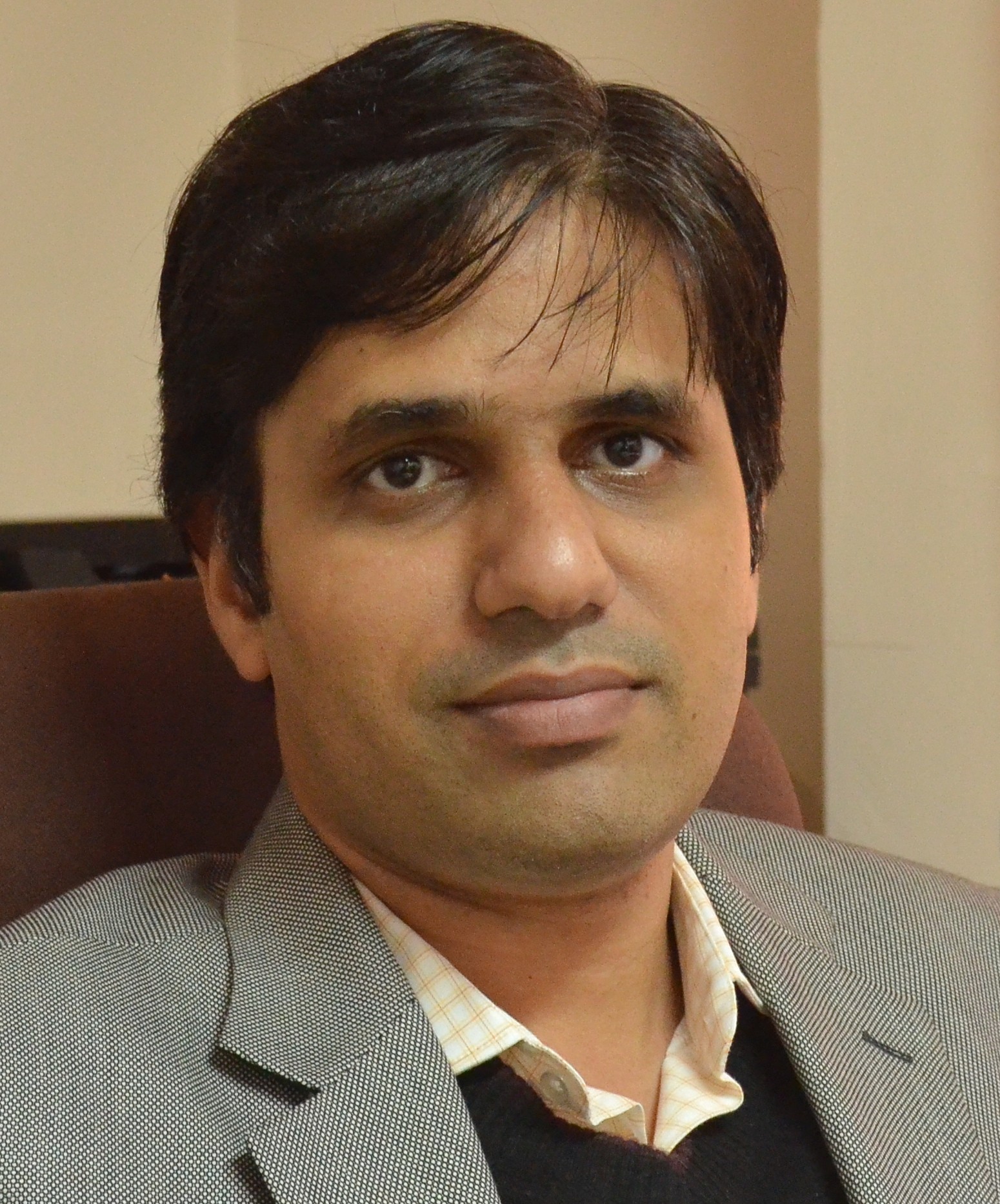}}]{Ketan Rajawat} (S'06--M'12) received his B.Tech and M.Tech degrees in Electrical Engineering from the Indian Institute of Technology (IIT) Kanpur, India, in 2007, and his Ph.D. degree in Electrical and Computer Engineering from the University of Minnesota, Minneapolis, MN, USA, in 2012. He is currently a Professor in the Department of Electrical Engineering, IIT Kanpur. His research interests are in the broad areas of signal processing, robotics, and communications networks, with particular
emphasis on distributed optimization and online learning. His current research focuses on the development and analysis of distributed and asynchronous optimization algorithms, online convex optimization algorithms, stochastic optimization algorithms, and the application of these algorithms to problems in machine learning, communications, signal processing, and smart grid systems. He is currently serving as a Senior Area Editor with the IEEE Transactions on Signal Processing. He is also the recipient of the 2018 INSA Medal for Young Scientists and the 2019 INAE Young Engineer Award.
\end{IEEEbiography}

\end{document}